\documentclass{aa}  
\usepackage{graphicx}
\usepackage{txfonts}
\usepackage{aalongtable}

\def\gtrsim{\mathrel{\hbox{\rlap{\hbox{\lower4pt\hbox{$\sim$}}}\hbox{$>$}}}}
\def\ltsim{\mathrel{\hbox{\rlap{\hbox{\lower4pt\hbox{$\sim$}}}\hbox{$<$}}}}

\def\kms{\hbox{{\rm km}\,{\rm s}$^{-1}$}}

\begin{document}
\title{Weak magnetic fields in Ap/Bp stars\thanks{Based on data obtained using the T\'elescope Bernard Lyot at Observatoire du Pic du Midi, CNRS and Universit\'e Paul Sabatier, France.}\fnmsep
\thanks{Fig. 7 to Fig. 32 are only available online. Table 3 is only available in electronic form at the CDS via anonymous ftp to cdsarc.u-strasbg.fr (130.79.128.5) or via http://cdsweb.u-strasbg.fr/cgi-bin/qcat?J/A+A/}}
\subtitle{Evidence for a dipole field lower limit and\\ a tentative interpretation of the magnetic dichotomy}

\author{M. Auri\`ere\inst{1}, G.A. Wade\inst{2}, J. Silvester\inst{2,3}, F. Ligni\`eres\inst{1}, S. Bagnulo\inst{4}, K. Bale \inst{2}, B. Dintrans\inst{1}, J.F. Donati\inst{1}, C.P. Folsom\inst{2}, M. Gruberbauer\inst{5}, A. Hui Bon Hoa\inst{1}, S. Jeffers \inst{6}, N. Johnson\inst{2}, J.D. Landstreet\inst{7}, A. L\`ebre\inst{8}, T. Lueftinger\inst{5}, S. Marsden\inst{9}, D. Mouillet\inst{1}, S. Naseri\inst{1}, F. Paletou\inst{1}, P. Petit\inst{1}, J. Power\inst{2}, F. Rincon\inst{1}, S. Strasser\inst{10}, N. Toqu\'e\inst{11}}

\offprints{M. Auri\`ere, {\tt michel.auriere@ast.obs-mip.fr}}

\institute{Laboratoire d'Astrophysique de Toulouse- Tarbes, Universit\'e Paul Sabatier, CNRS, Observatoire Midi-Pyr\'en\'ees, 57 Avenue d'Azereix, 65008 Tarbes, France
 \and
Department of Physics, Royal Military College of Canada,
  PO Box 17000, Station 'Forces', Kingston, Ontario, Canada K7K 4B4
\and
Department of Physics, Queen's University, Kingston, Ontario, Canada
\and
 Armagh Observatory, College Hill, Armagh BT61 9DG, Northern Ireland
\and
Institut f\"ur Astronomy, University of Vienna, T\"urkenschanzstrasse 17, A-1180 Wien, Austria
\and
Sterrekundig Instituut Utrecht, Universiteit Utrecht, PO Box 80000, 3508 TA Utrecht, The Netherlands
\and
 Department of Physics \& Astronomy, The University of Western Ontario, London, Ontario, Canada, N6A 3K7
\and
GRAAL, Universit\'e Montpellier II, CNRS, 62 Place Eug\`ene Bataillon, 34095 Montpellier, France
\and
Anglo-Australian Observatory, PO Box 296, Epping, NSW 1710, Australia
\and
Department of Astronomy, University of Minnesota, 116 Church Street, S.E., Minnepolis 55455, USA
\and
Department of Physics and Astronomy, Saint Mary's University, 923 Robie Street, Halifax, Nova Scotia, B3H3C3, Canada}

   \date{Received September 15, 1996; accepted March 16, 1997}


\abstract
{}{We have investigated a sample of 28 well-known spectroscopically-identified magnetic Ap/Bp stars, with weak, poorly-determined or previously undetected magnetic fields. The aim of this study is to explore the weak part of the magnetic field distribution of Ap/Bp stars.}
{Using the MuSiCoS and NARVAL spectropolarimeters at T\'elescope Bernard Lyot (Observatoire du Pic du Midi, France) and the cross-correlation technique Least Squares Deconvolution (LSD), we have obtained 282 LSD Stokes $V$ signatures of our 28 sample stars, in order to detect the magnetic field and to infer its longitudinal component with high precision (median $\sigma=40$~G). }
{For the 28 studied stars, we have obtained 27 detections of Stokes $V$ Zeeman signatures from the MuSiCoS observations. Detection of the Stokes $V$ signature of the $28^{\rm th}$ star (HD~32650) was obtained during science demonstration time of the new NARVAL spectropolarimeter at Pic du Midi.  This result shows clearly that when observed with sufficient precision, all firmly classified Ap/Bp stars show detectable surface magnetic fields. Furthermore, all detected magnetic fields correspond to longitudinal fields which are significantly greater than some tens of G. To better characterise the surface magnetic field intensities and geometries of the sample, we have phased the longitudinal field measurements of each star using new and previously-published rotational periods, and modeled them to infer the dipolar field intensity and the magnetic obliquity. The distribution of derived dipole strengths for these stars exhibits a plateau at about 1 kG, falling off to larger and smaller field strengths. Remarkably, in this sample of stars selected for their presumably weak magnetic fields, we find only 2 stars for which the derived dipole strength is weaker than 300~G. We interpret this ``magnetic threshold'' as a critical value necessary for the stability of large-scale magnetic fields, and develop a simple quantitative model that is able to approximately reproduce the observed threshold characteristics. This scenario leads to a natural explanation of the small fraction of intermediate-mass magnetic stars. It may also explain the near-absence of magnetic fields in more massive B and O-type stars.}{}{}
\authorrunning{M. Auri\`ere et al.}
\titlerunning{Magnetic fields of weak-field Ap stars}

\maketitle

\section{Introduction}

The magnetic chemically peculiar Ap/Bp stars are the non-degenerate stars for which the strongest magnetic fields have been measured (Landstreet 1992). Although the fields are thought to be fossil remnants of flux swept up during star formation or produced via dynamo action on the pre-main sequence, their origin is not understood in any real detail (e.g. Moss 2001). Furthermore, the role of the magnetic field in the diffusion processes which are responsible for their chemical peculiarity has been studied in only a schematic fashion.

Although more than one thousand main sequence {A-type stars} have been catalogued as magnetic Ap/Bp stars (Renson et al. 1991) following the scheme of Preston (1974), direct measurements of the magnetic field have been obtained for only a few hundred of them (Romanyuk 2000, Bychkov et al. 2003). Examination of the published measurements shows that the majority of the reported values are rather large. For example, $55\%$ of the 210 stars of the catalogue of Romanyuk (2000) with published magnetic field measurements have a maximum unsigned line-of-sight (longitudinal) magnetic field $B_\ell$ larger than 1 kG. On the other hand, according to Bohlender \& Landstreet (1990), the median root-mean-square (rms) longitudinal magnetic field of Ap stars (based on a small magnitude-limited sample observed by Borra \& Landstreet 1980) is only about 300~G (the largest rms field they report is only 710 G). This implies that most Ap stars have relatively weak ($\ltsim 1$~kG) magnetic fields, and that the available observations are strongly biased toward stars with the strongest and most easily-measured fields. One consequence of this bias is that the weak-field part of the magnetic field distribution of Ap stars is poorly studied. It is not known if it increases monotonically toward arbitrarily small field strength, or if it is truncated at a minimum magnetic field strength (as proposed by Glagolevskij \& Chountonov 2002). 

In order to improve our knowledge of the weak-field part of the magnetic field distribution of Ap stars, we have undertaken a study of a sample of 28 well-known spectroscopically-identified Ap/Bp stars, with very weak, poorly-determined or previously-undetected magnetic fields. We describe our survey in Sect. 2 and report our observational and modeling results in Sect. 3 and Sect. 4. We discuss the implications of our results, suggesting one possible interpretation involving the stability of large scale magnetic fields, in Sect. 5 and give our conclusions in Sect. 6.

\section {The weak-field Ap stars survey}

\subsection{The selected sample}

Our sample is composed of spectroscopically-identified Ap/Bp stars belonging to the HD catalogue. Twelve stars were selected based on the observations of Borra \& Landstreet (1980) and Bohlender et al. (1993), identifying stars for which no significant detection of the magnetic field was obtained. Thirteen additional targets are stars for which only old photographic measurements of the magnetic field were available, typically by Babcock (1958), and for which measurements have poor precision and do not provide a significant detection of the magnetic field; these stars were generally selected using the catalogues of Romanyuk (2000) or Bychkov et al. (2003). Finally, 3 stars of our sample were selected which had not been observed for magnetic field before this work. The observational properties of the 28 stars are presented in Table 1. Section 3.2 gives more details on each star and on the obtained results.

Because all of these stars are relatively bright, most have been known for decades and have been well studied. All appear in the Hipparcos catalogue (Perryman et al.,  1997), and all but 6 have $\sigma_\pi/\pi<0.2$. The majority have photometrically-determined rotational periods and published values of $v\sin i$. Many have been studied using high-resolution spectroscopy and Doppler Imaging. Therefore the classification of this sample as {\em bona fide} Ap/Bp stars is generally quite firm. As a consequence of this careful selection (and as will be described later in the paper), {\em no} non-Ap/Bp star has been mistakenly included in the sample.

\subsection{Observations and reduction}

Stokes $V$ and Stokes $I$ spectra of the 28 sample stars were obtained during 10 observing runs, from July 2001 to June 2006.  We used the MuSiCoS spectropolarimeter attached the Bernard Lyot telescope (TBL) at Observatoire du Pic du Midi. The MuSiCoS spectropolarimeter is composed of a cross-dispersed echelle spectrograph (Baudrand \& B\" ohm 1992) and a dedicated polarimeter module (Donati et al. 1999). The spectrograph is a table-top instrument, fed by a double optical fibre directly from the Cassegrain-mounted polarimeter. In one single exposure, this apparatus allows the acquisition of a stellar spectrum in a given polarisation (Stokes $V$ in this case) throughout the spectral range 450 to 660 nm with a resolving power of about 35000. Spectra in both orthogonal polarisations are recorded simultaneously by the CCD detector. A complete Stokes $V$ exposure consists of a sequence of four subexposures, between which the quarter-wave plate is rotated by 90$^{\rm o}$. This has the effect of exchanging the beams in the whole instrument, and in particular switching the positions of the two orthogonally polarised spectra on the CCD, thereby reducing spurious polarisation signatures. 

The echelle polarisation spectra were reduced using the ESpRIT package (Donati et al. 1997). The observation and reduction procedures are more thoroughly described by Shorlin et al. (2002). 

The correct operation of the MuSiCoS instrument, and in particular the absence of spurious magnetic field detections, is supported by other data obtained during these same observing runs, including studies of non-magnetic A-type stars (e.g. Shorlin et al. 2002), magnetic A, B and O-type stars (e.g., Ryabchikova et al. 2005a,  Donati et al. 2001, Wade et al. 2006a) and magnetic late-type stars (e.g. Petit et al. 2005).

MuSiCoS has recently been decomissioned, and has been replaced with NARVAL (Auri\`ere 2003), the new-generation spectropolarimeter which is a copy of the ESPaDOnS instrument in operation at the Canada-France-Hawaii Telescope (Donati 2004, 2007: in preparation). The main improvements of NARVAL in polarisation mode with respect to MuSiCoS are a spectral resolution of about 65000, spectral response between 370 nm and 1000 nm, and an overall sensitivity increased by a factor of about 30.

\begin{table*}[t]
\caption{Observational properties of the weak-field Ap star sample. Columns give ID and HD number, visual magnitude, spectral classification, effective temperature, luminosity and radius (with associated $1\sigma$ error bars), adopted LSD mask temperature, number of observations obtained and detection level (d=definite detection; m=marginal detection), maximum observed unsigned longitudinal field in G, $1\sigma$ error in G, and peak longitudinal field detection significance {\it z } = $ |B^{max}_{l}|$ / $\sigma$. $B_{\rm d}^{\rm min, 3.3}$ is the minimum dipole field (at $2\sigma$) inferred from the maximum measured longitudinal field and Eq. (7).}
\begin{tabular}{lcrr|ccc|cccccc}
\\
\hline
   ID & HD      & $m_V$ & Spec & $T_{\rm eff}$ & $\log L$ & $R$ & Mask &   \# & Det.& $|B_{\ell}|^{max}\pm \sigma$ & {\it z } & $B_{\rm d}^{\rm min, 3.3}$\\
    &      &  & Type &  (K) & ($L_\odot$) & ($R_\odot$) &  (kK) &  & level & (G)& &(G)\\
\hline
\\
HN And& 8441     &  6.7 &A2p  &$ 9060\pm 300$ & $ 1.90\pm	0.16$&$3.6	\pm	0.9 $   &  9  & 8& 8d       & $157\pm 18 $& 8.7 &  399    \\
43 Cas& 10221    &  5.5 &A0sp &$10660\pm 350$ & $ 2.11\pm	0.09$&$3.3	\pm	0.7 $   &  11  & 10& 8d     & $148\pm 34 $& 4.3 &  264    \\
$\iota$ Cas &15089 & 4.5 &A5p &$ 8360\pm 275$ & $ 1.38\pm	0.05$&$2.3	\pm	0.4 $   & 9  & 12& 12d      & $486\pm 23 $& 20.8&  1452   \\
&15144 & 5.8 &A6Vsp           &$ 8480\pm 280$ & $ 1.21\pm	0.07$&$1.9	\pm	0.3 $   & 9  & 6& 6d        & $631\pm 15 $& 42.0&  1983   \\
21 Per& 18296  & 5.0 &B9p     &$ 9360\pm 310$ & $ 2.08\pm	0.11$&$4.2	\pm	1.0 $   & 10  & 2& 2d       & $213\pm 20 $& 10.6&  571    \\
9 Tau & 22374 & 6.7 &A2p      &$8390\pm  275$ & $ 1.48\pm	0.13$&$2.6	\pm	0.7 $   &  9  & 2& 2d       & $523\pm 24 $& 21.7&  1568   \\
56 Tau &27309    &  5.3 &A0sp &$12730\pm 420$ & $ 2.06\pm	0.08$&$2.2	\pm	0.4 $   & 12  & 12& 12d     & $804\pm 50 $& 16.1&  2323   \\  
11 Ori& 32549    &  4.7 &A0sp &$10220\pm 335$ & $ 2.35\pm	0.12$&$4.7	\pm	1.2 $   & 11  & 11 & 1d3m &  $186 \pm 39$& 4.7 &  356    \\  
&32650 & 5.4 &B9sp            &$11920\pm 390$ & $ 2.11\pm	0.07$&$2.7	\pm	0.5 $   & 12  & 18 & 2m3d&   $91  \pm 18$& 5.0 &  237     \\
&37687 & 7.0 &B8              &$9450\pm  310$ & $ 2.18\pm	0.25$&$4.6	\pm	1.9 $  & 10  & 2 & 2d       & $766\pm 119$& 6.4 &  1742   \\
137 Tau& 39317 & 5.5 &B9spe   &$10130\pm 330$ & $ 2.19\pm	0.14$&$4.0	\pm	1.1 $   & 11  & 8 & 3d1m  &  $216 \pm 59$& 3.6 &  323    \\
&40711 & 8.5 &A0              &$8070\pm  265$ & $ 1.94\pm	0.61$&$4.8	\pm	5.3 $   &  9  & 3 & 1d1m  &  $528 \pm 38$& 13.8&  1492   \\
&43819 & 6.2 &B9IIIsp         &$10880\pm 355$ & $ 2.15\pm	0.20$&$3.3	\pm	1.2 $   &  11  & 8 & 8d     & $628\pm 25 $& 25.1&  1907   \\
15 Cnc& 68351    &  5.6 &B9sp &$10290\pm 340$ & $ 2.65\pm	0.21$&$6.6	\pm	2.4 $   &  10  & 16& 1d4m &  $325 \pm 47$& 6.9 &  762    \\
3 Hya &72968    &  5.7 &A1spe &$9840\pm  320$ & $ 1.55\pm	0.08$&$2.0	\pm	0.4 $   & 10  & 13& 13d     & $427\pm 16 $& 27.4&  1304   \\
45 Leo& 90569    &  6.0 &A0sp &$10250\pm 335$ & $ 1.78\pm	0.10$&$2.5	\pm	0.5 $   & 11  & 10& 10d     & $541\pm 23 $& 23.5&  1634   \\ 
&94427    &  7.3 &A5          &$7250\pm  240$ & $ 1.05\pm	0.11$&$2.1	\pm	0.5 $   & 8  & 8& 8d        & $356\pm 41 $& 8.6 &  904    \\  
EP Uma &96707 & 6.0 &F0sp     &$7780\pm  255$ & $ 1.54\pm	0.08$&$3.2	\pm	0.6 $   & 8  & 21& 7d7m   &  $69  \pm 33$& 2.3 &  128     \\
65 Uma& 103498 & 6.9 &A1spe   &$9220\pm  300$ & $ 2.06\pm	0.20$&$4.2	\pm	1.5 $   & 9  & 14& 12d1m  &  $169 \pm 19$& 8.9 &  432    \\
21 Com& 108945   &  5.4&A2pvar&$8870\pm  290$ & $ 1.72\pm	0.09$&$3.1	\pm	0.6 $   & 9  & 13& 12d1m  &  $234 \pm 54$& 4.3 &  416    \\ 
$\omega$ Her &  148112&4.6&B9p&$9330\pm  305$ & $ 1.86\pm	0.08$&$3.2	\pm	0.6 $   & 10  & 12& 11d     & $204\pm 21 $& 9.7 &  535    \\ 
45 Her &151525   &  5.2 &B9p  &$9380\pm  310$ & $ 2.18\pm	0.13$&$4.7	\pm	1.2 $   & 11  & 14 & 2d3m &  $146 \pm 38$& 3.8 &  231    \\
&171586 & 6.4&A2pvar          &$8760\pm  290$ & $ 1.37\pm	0.10$&$2.1	\pm	0.5 $   & 10  & 5 & 5d      & $375\pm 56 $& 6.6 &  868    \\
&171782 & 7.8 &A0p            &$9660\pm  315$ & $ 1.76\pm	0.30$&$2.7	\pm	1.3 $  & 10  & 6 & 2d1m   &  $333 \pm 78$& 4.2 &  584    \\
19 Lyr &179527   &  5.9 &B9sp &$10370\pm 340$ & $ 2.63\pm	0.16$&$6.4	\pm	1.9 $   &   11  & 11& 8d2m&  $156 \pm 46$& 3.4 &  211    \\
4 Cyg& 183056   &  5.1 &B9sp  &$11710\pm 385$ & $ 2.69\pm	0.11$&$5.3	\pm	1.2 $   & 12  & 13& 13d     & $290\pm 42 $& 6.9 &  680    \\
&204411   &  5.3 &A6pe        &$8750\pm  290$ & $ 1.97\pm	0.07$&$4.2	\pm	0.8 $   &   9  & 12& 12d    & $88 \pm 14 $& 6.0 &  198    \\
$\kappa$ Psc &220825  &4.9&A0p&$9450\pm  310$ & $ 1.40\pm	0.05$&$1.9	\pm	0.3 $   &   10  & 12& 12d   & $312\pm 25 $& 12.8&  865    \\
\hline

\end{tabular}
\end{table*}

\subsection{Physical properties of the sample}

For each of the sample stars, we have determined effective temperature $T_{\rm eff}$, luminosity $L$ and radius $R$, to allow us to identify the appropriate line mask for Least-Squares Deconvolution (Sect. 2.5) and for determination of the rotational axis inclination for the dipole magnetic field model (Sect. 2.7).

Effective temperatures of stars of our sample were derived using Geneva and Str\"omgren photometry (obtained from the General Catalogue of Photometric Data (GCPD); Mermilliod et al. 1997) using the calibrations of Hauck \& North (1982) and Moon \& Dworetsky (1985). Effective temperatures reported in Table 1 are the average of the two estimates when both were available, or the single one which could be derived when only one photometric set was available. We have assumed for $T_{\rm eff}$ an uncertainty of the order of 3\% for propagation of uncertainties in all calculations using the effective temperature.

Luminosity was inferred using the GCPD-reported visual magnitude, the Hipparcos parallax and the bolometric correction relations of Balona (1994). Radius was then inferred directly from the luminosity and temperature via the Stefan-Boltzmann equation for a uniform spherical star.

The inferred values of $T_{\rm eff}$, $\log L/L_\odot$ and $R/R_\odot$ are reported in Table 1. For the 20 stars which we have in common with the study of Kochukhov \&  Bagnulo (2006), these values are all in good agreement. As pointed out by Landstreet et al. (2007), the assumed uncertainties of  Kochukhov \& Bagnulo (2006), which are comparable to our own, are probably somewhat underestimated. However, as our fundamental parameters are not to be used for detailed evolutionary studies, we consider them to be sufficient for this study.

\subsection{Least-Squares Deconvolution and magnetic field detection}

 The primary aim of our study is to detect line circular polarisation (a ``Stokes $V$ Zeeman signature'') which is characteristic of the longitudinal Zeeman effect produced by the presence of a magnetic field in the stellar photosphere. For this we used the Least-Squares Deconvolution (LSD) procedure, first used by Donati et al. (1997) to study the magnetic fields of active late-type stars and by Wade et al. (2000 a,b) for Ap stars. This method enables the ``averaging'' of several hundred (and possibly several thousand in some stars) lines and thus to obtain Stokes $I$ and Stokes $V$ profiles with greatly improved S/N. 

LSD provides a single quantitative criterion for the detection of Stokes $V$ Zeeman signatures: we perform a statistical test in which the reduced $\chi^2$ statistic is computed for the Stokes $V$ profile, both inside and outside the spectral line (Donati et al. 1997). The statistics are then converted into detection probabilities, which are assessed to determine if we have a definite detection (dd, false alarm probability smaller than $10^{-5}$), a marginal detection (md, false alarm probability greater than $10^{-5}$ and smaller than $10^{-3}$), or no detection at all (nd). A diagnostic null spectrum (called $N$ in the following) is also obtained using the same subexposures obtained for Stokes $V$, but by pair processing those corresponding to identical azimuths of the quarter-wave plate. By checking that a signal is detected only in $V$ and not in $N$, and that any detected signature is located within the line profile velocity interval, we can distinguish between real magnetic signatures and (infrequent) spurious signatures. In addition, using the full resolved Stokes $V$ profile enables the detection of the magnetic field, even if the integrated line-of-sight component is very weak, or even null.

\subsection {LSD masks}

LSD is a cross-correlation method which requires comparison of our observed spectra with synthetic line masks (Donati et al. 1997, Shorlin et al. 2002). To obtain the most realistic masks, we used spectral line lists from the Vienna Atomic Line Database (VALD; Piskunov et al 1995; Ryabchikova et al. 1997; Kupka et al. 1999). To take into account the chemical peculiarities of Ap/Bp stars, we employed an abundance table in which the abundance of metals (Al, Si, S, Ti, V, Mn, Fe, Co, Ni, Zn, Sr, Y, Zr, Ba, La, Ce, Pr, Nd, Eu, Gd, Dy) is 10x solar, except for Cr which was increased to 100x solar (e.g. Shorlin et al. 2002). Masks were then compiled for effective temperature ranging from 7000-13000~K, with $\log g=4.0$, a microturbulence of 2~km/s, and including all metal lines with a central depth greater than 10\% of the continuum. We also compiled a series of masks assuming solar abundances. We have found in other studies that the magnetic field measurements are not very sensitive to the mask temperature within a couple of thousand K (Wade et al., in preparation). We therefore computed masks spaced every 1000~K. 

Least-Squares Deconvolution was performed for several temperatures and in some cases the most significant magnetic field detection was obtained for a temperature somewhat hotter than that given by the photometric data. This occured several times for the hottest sample stars - this can be seen in Table 1. In the case of the cool Ap star EP UMa, a solar abundance mask gave a better result (better detection of magnetic field and smaller error bars on $B_\ell$) than Ap abundances as described above. These discrepancies probably result from differences between the true chemical peculiarities of individual stars and those assumed in the line masks. For the discrepant hot stars, we computed the longitudinal magnetic field for a mask temperature between that corresponding to the best detection and the derived effective temperature. For EP UMa, we used solar abundance masks in our analysis. The number of lines used in the LSD ranged from 1500 to 3000, and is anticorrelated with the temperature. 

\subsection{Longitudinal magnetic field}

The longitudinal magnetic field was inferred from each of the Stokes $I$ and $V$ profile sets, using the first-order moment method. According to this method, the longitudinal field $B_\ell$ (in G) is calculated from the Stokes $I$ and $V$ profiles in velocity units as:

\begin{equation}
B_{\ell} = -2.14 \times 10^{11} \frac{\int_{}^{}vV(v)dv}{\lambda gc\int_{}^{}[I_c - I(v)]dv},
\end{equation}

\noindent (Rees \& Semel 1979; Donati et al. 1997; Wade et al. 2000b) where $\lambda$, in
nm, is the mean wavelength of the LSD profile, $c$ is the velocity of light (in the same units as $v$), and $g$ is the mean
 value of the Land\'e factors of all
lines used to construct the LSD profile. Integration ranges used for evaluation of Eq. (1) were computed automatically, beginning and ending 15~km/s before/after the location in the line wings at which the residual flux was equal to 85\% of the continuum flux. The accuracy of this
technique for determining high-precision longitudinal field
measurements has been clearly demonstrated by Wade et al. (2000b),
Donati et al. (2001) and Shorlin et al. (2002).

 The resultant longitudinal magnetic field measurements, which are reported in Table~3, are remarkably precise (this Table is only available on line). The 282 measurements, with a median $1\sigma$ uncertainty of 40 G, represent the largest compilation of high-precision stellar magnetic field measurements ever published.

\subsection{Modeling the longitudinal field variation}

To characterise the dipole components of the magnetic fields of our sample stars, we use the oblique rotator model (ORM, Stibbs 1950) as formulated by Preston (1967). This model provides a good first approximation of the large scale magnetic field of Ap stars (e.g. Landstreet 1988). Because of the weakness of the longitudinal magnetic field observed for the stars of our sample, we do not expect to be able to detect departures from a global dipolar configuration. 

To begin, we added to our high-precision data set additional good-quality published magnetic field measurements collected by Bychkov et al. (2003). Details of these collected measurements were kindly provided by Dr. Victor Bychkov. Then, we searched the literature for rotational periods for each of our sample stars. For many stars, published rotational periods were available which provided an acceptable folded phase variation $B_\ell(\phi)$ of the magnetic measurements. However, for some stars, the published rotational period or periods did not provide an acceptable folded magnetic field variation, and for others, no published period was available. For these latter stars, we used a modified Lomb-Scargle technique to attempt to infer the rotational period, both directly from the longitudinal field measurements, as well as from the variations of the LSD Stokes $I$ and $V$ profiles. The period searches of LSD profiles were performed by treating each pixel in the Stokes $I$ and $V$ profiles as an independent timeseries (similar to the technique described by Adelman et al., 2002). Individual periodograms were subsequently weighted according to their amplitude of variation and averaged to characterise variability of the whole LSD profile. Acceptable periods were identified by establishing the 99\% confidence threshold, and candidate periods were evaluated by phasing the LSD profiles and longitudinal field measurements.  

Results of the period searches for individual stars are provided in their appropriate subsections in Sect. 3. Ultimately, acceptable rotational periods were obtained for 24 stars, and these periods are reported in Table 2.

We also searched the literature for published values of the projected rotational velocity ($v\sin i$) of each star, which we compared to the value measured from the LSD Stokes $I$ profile by fitting rotationally-broadened synthetic profiles. Sometimes significant discrepancies were found between our values of $v\sin i$ and those reported in the literature. These discrepacies are discussed in Sect. 3, and the adopted rotational velocities (generally those obtained from the LSD profiles) are shown in Table 2. 

Each phased longitudinal field variation $B_\ell(\phi)$ was then fit using a 1$^{\rm st}$ order sine function:

\begin{equation}
B_\ell(\phi) = B_0 + B_1\sin {2\pi(\phi+\phi_0)}.
\end{equation}


The phased and fit longitudinal field variations are shown in Fig. 1. The reduced $\chi^2$ of this fit ($\chi^2_2$), along with those of linear fits through $B_\ell=0$ (the ``null field'' model, $\chi^2_0$ ) and through the weighted mean of the measurements (the ``constant field'' model, $\chi^2_1$),  of each star, are reported in Table 2.  $\chi^2_{\rm lim}$,  the $2\sigma$ upper limit for admissible models, computed according to Press et al. (1992), is reported in Table 2 as well . A comparison of these reduced $\chi^2$ values with each other allows us to evaluate the significance of the detection of the longitudinal magnetic field and its variability. Although variability of the longitudinal field cannot be established for a few stars ($\chi^2_1<\chi^2_{\rm lim}$), the only star for which the longitudinal field is not detected with more than 2$\sigma$ confidence is HD 96707 (for which $\chi^2_0<\chi^2_{\rm lim}$).

\begin{table*}[ht]
\caption{Results of the magnetic field modeling. The contents of the columns are described in sect. 2.7. The uncertainties associated with the derived dipole parameters $i, \beta$ and $B_{\rm d}$ correspond to 2$\sigma$.}
\begin{tabular}{lcc|cccc|cccccc}
\\
\hline
star     &   Period   &  $v\sin i$& $\chi^2_0$ & $\chi^2_1$ &$\chi^2_2$ &$\chi^2_{\rm lim}$& $i$ &$\beta$& $B_{\rm d}$& $B_{\rm d}^{\rm min}$& $B_{\rm d}^{\rm max}$\\
         & (d)             & (km/s)           &            &            &           &                  & ($\degr$) & ($\degr$) & ($10^3$ G)       & ($10^3$ G)                  & ($10^3$ G)\\    
\hline
8441  & 69.2         & 2  &    13.71 &     6.69 &     1.75 &3.35 &$  49\pm33$ & $ 73_{- 61}^{+ 17}$ &      0.683 &      0.415 &     2.931 \\
10221 & 3.15459      & 24 &     5.90 &     1.81 &     1.56 &2.71 &$  27\pm11$ & $ 42_{- 41}^{+ 38}$ &      0.375 &      0.195 &     1.202 \\ 
15089 & 1.74033      & 48 &    57.83 &    56.37 &     1.83 &2.14 &$  45\pm11$ & $ 80_{- 12}^{+  7}$ &     2.031 &     1.560 &     2.999 \\
15144  & 2.99787     & 13 &  1305.83 &     4.57 &     0.04 &2.71 &$  24\pm 8$ & $  9_{-  3}^{+  6}$ &     2.100 &     2.007 &     2.281 \\ 
27309 & 1.568884     & 57 &   157.42 &     2.43 &     1.74 &2.63 &$  53\pm17$ & $  5_{-  5}^{+ 11}$ &     3.673 &     2.325 &     8.022 \\  
32549 & 4.6393       & 47 &     7.19 &     8.56 &     1.75 &2.75 &$  65\pm27$ & $ 77_{- 74}^{+ 11}$ &      0.546 &      0.312 &    28.176 \\
32650 & 2.7347       & 30 &     4.48 &     1.92 &     1.19 &1.72 &$  37\pm12$ & $ 45_{- 30}^{+ 31}$ &      0.229 &      0.153 &      0.477 \\
39317 & 2.6541       & 45 &     4.97 &     1.84 &     2.24 &3.18 &$  36\pm12$ & $ 20_{- 20}^{+ 69}$ &      0.560 &      0.113 &     2.252 \\    
43819 & 15.02        & 10 &   135.66 &    59.32 &     3.45 &4.18 &$  63\pm66$ & $ 42_{- 42}^{+ 47}$ &     2.626 &     2.488 &    78.367 \\   
68351 & 4.16         & 33 &     7.85 &     2.17 &     1.38 &2.00 &$  28\pm37$ & $ 46_{- 41}^{+ 36}$ &      0.649 &      0.437 &    71.486 \\    
72968 & 5.6525       & 16 &   360.16 &     4.51 &     1.22 &2.03 &$  61\pm18$ & $  5_{-  4}^{+  7}$ &     2.388 &     1.451 &     6.702 \\   
90569 & 1.04404      & 13 &   170.38 &    36.08 &     1.93 &3.07 &$   9\pm 4$ & $ 81_{-  7}^{+  5}$ &     5.157 &     2.946 &    11.284 \\     
94427  & 1.9625      &  8 &    37.58 &    46.61 &     2.12 &3.72 &$   8\pm 4$ & $ 89_{-  4}^{+  1}$ &     8.957 &     3.806 &    25.519 \\    
96707  & 3.515       & 37 &     0.82 &     0.91 &     0.77 &1.21 &$  53\pm16$ & $ 90_{- 90}^{+  0}$ &      0.100 &        0.0 &      0.492 \\    
103498 & 15.830      & 13 &    25.43 &    29.71 &     6.55 &7.44 &$  75\pm68$ & $ 80_{- 11}^{+ 10}$ &      0.600 &      0.572 &     6.751 \\    
108945 & 2.01011     & 65 &     5.31 &     6.00 &     2.10 &2.83 &$  57\pm18$ & $ 85_{- 61}^{+  3}$ &      0.735 &      0.333 &     1.509 \\    
148112 & 3.04296     & 44.5 &  40.67 &     0.93 &     0.95 &1.73 &$  56\pm16$ & $  3_{-  3}^{+ 11}$ &     1.042 &      0.579 &     2.370 \\  
151525 & 4.1164      & 35 &     2.74 &     2.86 &     0.97 &1.70 &$  37\pm19$ & $ 78_{- 43}^{+ 11}$ &      0.545 &      0.208 &     1.927 \\       
171586 & 2.1308      & 37 &    10.86 &     8.27 &     2.59 &6.60 &$  48\pm19$ & $ 46_{- 40}^{+ 40}$ &     1.422 &      0.716 &     4.413 \\    
171782 & 4.4674      & 24 &     6.88 &     1.12 &     1.79 &3.79 &$  51\pm51$ & $  5_{-  5}^{+ 85}$ &     1.651 &      0.213 & 22257.276 \\    
179527 & 7.098       & 33 &     6.57 &     7.70 &     0.30 &1.30 &$  74\pm32$ & $ 81_{- 34}^{+  8}$ &      0.522 &      0.409 &     1.233 \\    
183056 & 2.9919      & 26 &    25.02 &    28.74 &     1.59 &2.32 &$  74\pm 8$ & $ 49_{- 32}^{+ 37}$ &     1.558 &     1.172 &     3.938 \\    
204411 & 4.8456      & 5.4&    20.75 &     5.43 &     0.57 &1.46 &$   7\pm 5$ & $ 81_{- 12}^{+  7}$ &      0.968 &      0.416 &     4.509 \\    
220825 & 1.42539     & 38 &    78.54 &    90.05 &     2.29 &3.18 &$  35\pm54$ & $ 83_{- 80}^{+  7}$ &     1.957 &     1.141 &    21.045 \\    
\\
\hline
\end{tabular}
\end{table*}

For a tilted, centred magnetic dipole, the surface polar field strength $B_{\rm d}$ is derived from the variation of the longitudinal magnetic field $B_\ell$ with rotational phase $\phi$ using Preston's (1967) well-known relation:

\begin{equation}
B_d = B_\ell^{\rm max} \biggl({{15 + u}\over{20(3-u)}} (\cos\beta\cos i + \sin\beta\sin i)\biggr)^{-1},
\end{equation}

\noindent where $B_\ell^{\rm max}=|B_0|+B_1$ and $u$ denotes the limb darkening parameter 
(equal to approximately $u=0.5$ for our sample).  The rotational axis inclination and
obliquity angles $i$ and $\beta$ are related by

\begin{equation}
\tan\beta={{1-r}\over{1+r}} \cot i,
\end{equation}

\noindent where $r=(|B_0|-B_1)/(|B_0|+B_1)$. 

We have determined the inclination $i$ for each of our stars assuming rigid rotation, and computing:

\begin{equation}
\sin i = {{P_{\rm rot} v\sin i}\over {50.6 R}},
\end{equation}

\noindent where $P_{\rm rot}$ is the adopted stellar rotational period in days, $v\sin i$ is the adopted projected rotational velocity in km/s, and $R$ is the computed stellar radius in solar units. The magnetic obliquity $\beta$ was then inferred from Eq. (4) and the polar strength of the dipole $B_{\rm d}$ from Eq. (3). Uncertainties associated with all parameters (at 2$\sigma$, including a max$(2~{\rm km/s}, 10\%)$ uncertainty on $v\sin i$) were propagated through the calculations of $i$, $\beta$ and $B_{\rm d}$. The resultant dipole magnetic field models are reported in Table 2. 

An important and interesting consequence of Eq. (3), independent of all parameters except limb-darkening, is:

\begin{equation}
B_{\rm d} \geq {{20(3-u)}\over{15+u}} B_\ell^{\rm max},
\end{equation}

\noindent which, for typical limb-darkening $u$, yields:

\begin{equation}
B_{\rm d} \gtrsim 3.3 B_\ell^{\rm max}.
\end{equation}

Therefore, exclusive of the model geometry, a lower limit on the surface dipole component of the magnetic field is obtained directly from the maximum measured value of the longitudinal field. The maximum measured longitudinal field is reported in column 11 of Table 1, and the inferred lower limit ($2\sigma$) of $B_{\rm d}$,  $B_{\rm d}^{\rm min, 3.3}$ is reported in column 13 of Table 1. It is clear from these data that $B_{\rm d}$ for most of our sample is larger than a few hundred G at $2\sigma$.

\section{Observational results}

\subsection {Global results as compared to previous work}

All of the 28 Ap stars studied were found to exhibit significant circular polarisation in their spectral lines. Table 1 gives our global result. For a majority of the sample, the detection of the magnetic field was obtained during the first observation. For some objects, obtaining a positive detection of the magnetic field required several observations, sometimes spanning several observing seasons. This was due to observations at phases in which the polarisation was especially weak, and  due to low S/N ratio resulting from poor meteorological conditions. The number of observations and of Zeeman detections of each star is also indicated in Table 1.

No significant or recent detection of the magnetic field has been reported in the past for essentially all stars of our sample. Shorlin et al. (2002), using MuSiCoS and the same procedure as in this paper, did not detect the magnetic field of 3 Ap stars they observed. For two of these stars, they suggested a misclassification. We re-observed the third star, HD 148112 ($\omega$ Her), and easily detected its field. Glagolevskij \& Chountonov (2002) observed 11 weakly-magnetic Ap stars and detected no magnetic fields in any of them. More recently, Glagolevskij et al. (2005) presented 9 stars in which they detected no longitudinal fields, with $B_\ell$ weaker than 100~G. Five of these stars were observed during this work and the magnetic field was detected for all of them. Hubrig et al. (2006) and Kochukhov \&  Bagnulo (2006) reduced FORS1 observations of about 100 Ap stars, among which were about 50 stars for which they detected the magnetic field for the first time. Four stars of their sample are in common with ours: 9 Tau, 21 Com, $\omega$ Her and 45 Her. In this study, the magnetic field is detected and measured for each of them, we give details in Sect. 3.2. on the results of the FORS1 investigations.

Our higher detection rate of magnetic field  with respect to other techniques is certainly due to in part to the S/N improvement associated with the LSD method. In addition, our magnetic field detection criterion is based on measurement of significant circular polarisation within the velocity-resolved line profile (the Stokes $V$ signature). This signature amplitude is only weakly sensitive to the global magnetic field geometry, and generally varies by only a factor of a few during the stellar rotational cycle, unlike the longitudinal field (which frequently varies by orders of magnitude, and which can vanish completely at some phases). Finally, the fact that we insist on performing multiple observations of each star is also an important factor in our superior magnetic field detection rate.

One set of LSD profiles will be shown for each of the studied stars in the online version of the paper. However, Figs. 2 and 3 illustrate the case of 21 Per (HD 18296) when the field was respectively $213\pm 20$~G ($10\sigma$ longitudinal field detection)
and $-39\pm 24$~G (insignificant longitudinal field detection). However, both of these observations correspond to highly significant detections of the Stokes $V$ Zeeman signatures.  Fig. 4 illustrates the non-detection of magnetic field in HD 32650 with MuSiCoS, and Fig. 5 shows the Zeeman detection of this same star obtained with NARVAL on 12 March 2007. 

We now present our Zeeman detection and modeling results for each star of our sample.

 \begin{figure*}[t]
   \centering
   \includegraphics[width=13cm, angle=-90]{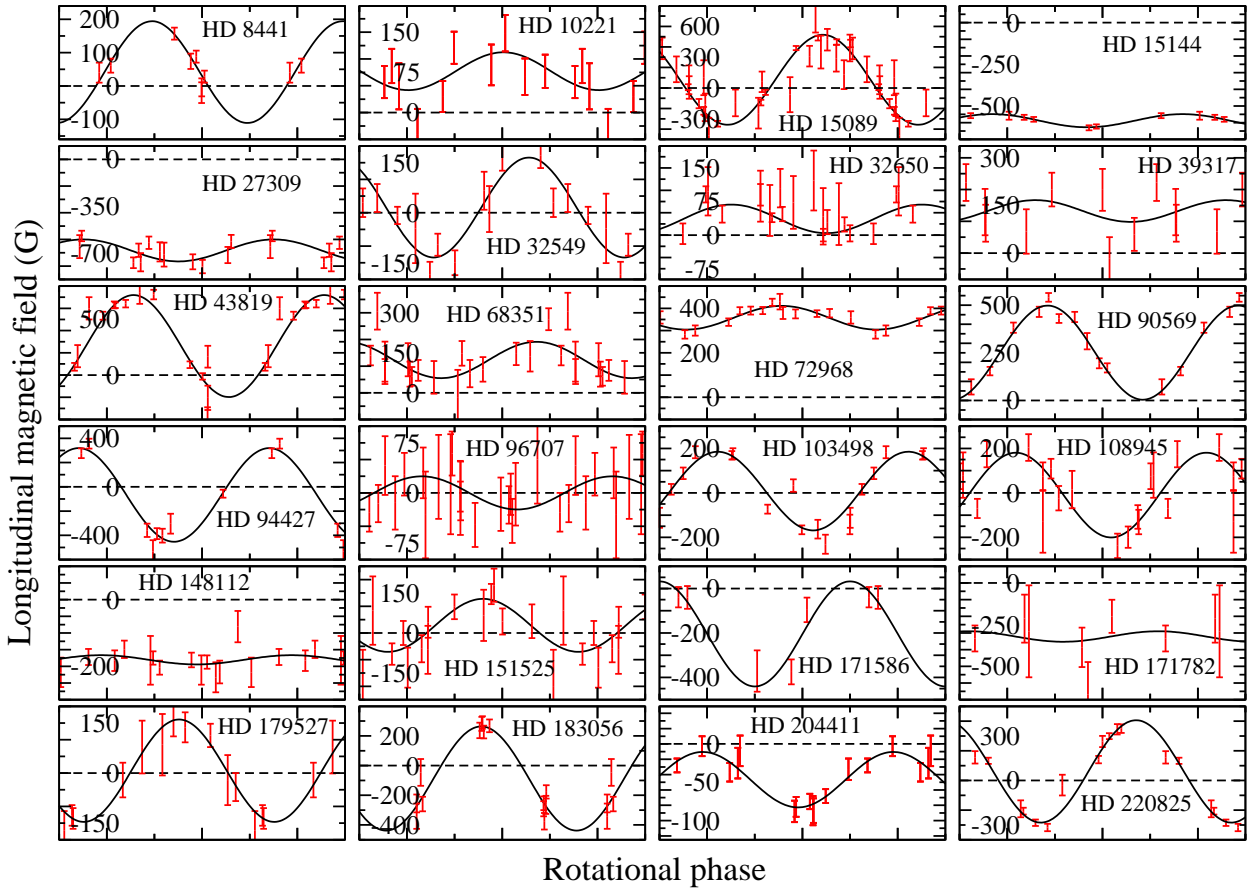}
           \label{}
\caption{Phased longitudinal field measurements of sample stars. The vertical (longitudinal field) scale differs in each frame, and is indicated in each frame in units of G. The horizontal (phase) axis in each frame runs from -0.25 to 1.25 cycles. }
   \end{figure*}

 \begin{figure}[t]
   \centering
   \includegraphics[width=7cm, angle=0]{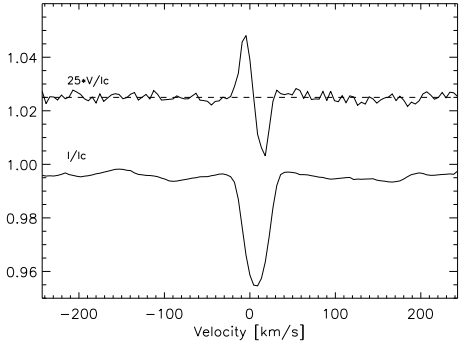}
         \label{}
\caption{LSD profiles of 21 Per (HD  18296) on 01 Sep 05.  From bottom to top, Stokes $I$ and Stokes $V$ are presented. For display purposes, the profiles are shifted vertically, and  Stokes $V$ profile is expanded by a factor of 25. The dashed line illustrates the zero level for the Stokes $V$ profile.}
   \end{figure}

 \begin{figure}[t]
   \centering
  \includegraphics[width=7cm, angle=0]{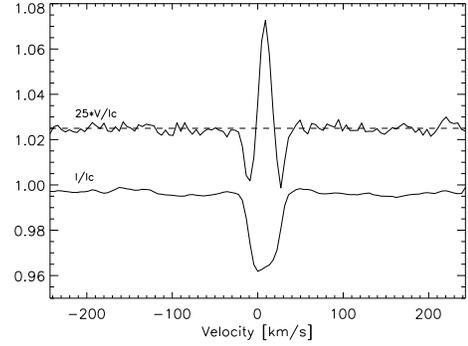}
         \label{}
\caption{LSD profiles of 21 Per on 02 sep 05 (same as Fig. 2).}
  \end{figure}

\begin{figure}[t]
 \centering
  \includegraphics[width=7cm, angle=0]{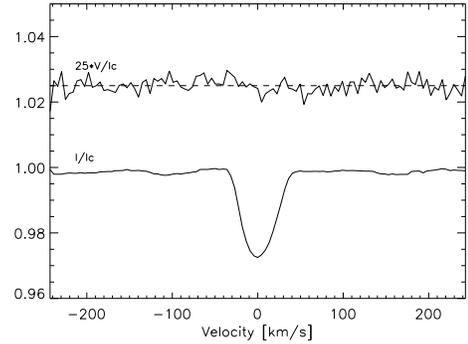}
        \label{}
\caption{LSD profiles (ND) of HD 32650 on 12 feb 04 observed with MuSiCoS (same as Fig. 2, no Zeeman detection).}
  \end{figure}

 \begin{figure}[t]
   \centering
   \includegraphics[width=7cm, angle=0]{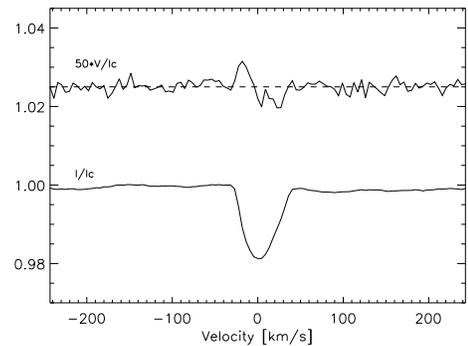}
         \label{}
\caption{LSD profiles (DD) of HD 32650 on 12 Mar 07 as observed with NARVAL (same as Fig. 2 but Stokes $V$ profile is expanded by a factor of 50, definite Zeeman detection).}
   \end{figure}

\subsection {Results obtained for each star}

\subsubsection {HN And (HD 8441)}

HN And is a well-studied SB1 A2p star (e.g. North et al. 1998, Adelman et al. 1995) for which no clear magnetic field detection has been reported. Babcock (1958) presented measurements of a relatively weak (some hundreds of G) field, and Bychkov et al. (2003) did not present any new significant Zeeman detection. Preston (1971) and Mathys \& Lanz (1992) did not succeed in measuring its field. We observed this star 8 times and obtained 8 definite Zeeman detections. Fig. 7 shows the LSD Stokes $V$ and Stokes $I$ profiles when the longitudinal magnetic field was $B_\ell = 155\pm 18$~G.

In the literature we find periods for HN And of 2.9632 days (Steinitz, 1965) based on magnetic data from Babcock 
(1958), 106.27 days from radial velocity measurements
(Renson, 1965) later shown 
to be the rotation period of the binary (North et al, 1998), 69.5 days 
based on Stromgren (Wolff \& Morrison, 1973) and Johnson (Rakosch \& Fiedler, 1978) 
photometry, and 1.80889 days (Bychkov et al, 2005) 
based mostly on the same magnetic data as Steinitz (1965).  Examining our 
periodogram near these periods, we only find a strong $\chi^2$ minimum near 69.5 days 
in both Stokes $V$ and $I$. The LSD profiles phase in a sensible fashion at this minimum, 
thus we can be confident that it is the correct period. Based on our analysis we adopt $P_{\rm rot}=69.2^{+0.6}_{-0.5}$~d,
where the uncertainties correspond to 99\% confidence.

The projected rotational velocity of HN And is very low, and is also not well known. North et al. (1998) report $v\sin i<2.35$~km/s. Our $R=35000$ spectra do not allow us to improve upon this estimate. Using a high-resolution ($R=110,000$) CFHT-Gecko spectrum of this star in our possession, we have attempted to estimate $v\sin i$ by modeling the 6150~\AA\ region. The best fit is obtained for $v\sin i \simeq 3$~km/s and very weak magnetic field ($0-500$~G). However, acceptable fits are also obtained for moderate fields (1-1.5~kG) and intermediate rotation (1.5-2~km/s). For calculation of the dipole field, we adopt $v\sin i=2$~km/s which results (at $2\sigma$ confidence) in $i=49\pm33$, $\beta=73^{+17}_{-61}$ and $415\leq B_{\rm d}\leq 2931$~G (with $B_{\rm d}^{\rm min, 3,3}=399$~G). Larger values of $v\sin i$ result in an increased value of the inferred $B_{\rm d}$, and assuming $v\sin i=1$~\kms (which we believe is the smallest value admissible by the line profiles), the weakest admissible field remains essentially unchanged.

\subsubsection {43 Cas, HD 10221}

43 Cas is a classical Ap star in which magnetic field has been sought for around 50 years, but neither Babcock (1958), Borra \& Landstreet (1980) nor Glagolevskij \& Chountonov (2002) could detect it. Glagolevskij et al. (2005) reported a $4 \sigma$ detection. In this paper, the Stokes $V$ Zeeman signal is clearly detected on 8 occasions.  Fig. 8 shows the LSD Stokes $V$ and Stokes $I$ profiles when the longitudinal magnetic field was measured to be $B_\ell = 148\pm 34$~G. Hildebrandt (1975) found a photometric period of 3.1848 d. for 43 Cas, which was refined to 3.15459 d from Hipparcos data (van Leeuwen et al. 1997). The projected rotational velocity, measured from modeling of the strongly-variable LSD Stokes $I$ profiles, is $24$~km/s. We derive  $i=27 \pm11\degr$,  $\beta=42^{+38}_{-41}\degr$ and $195\leq B_{\rm d}\leq 1202$~G (with $B_{\rm d}^{\rm min, 3,3}=264$~G). 

\subsubsection {$\iota$ Cas, HD 15089}

The magnetic field of $\iota$ Cas was not detected by Borra \& Landstreet (1980), but was by Kuschnig et al. (1998). As shown in Fig. 9 and Table 3, the field is easily detected (12 definite detections), and a rather large longitudinal field is measured ($B_\ell = - 348\pm 28$~G). The rotational period of this star has been recently confirmed by Hipparcos (1.74033 d), and Kuschnig et al. (1998) give $v\sin i=48$~km/s (consistent with Royer et al. 2002, and somewhat lower than the LSD profiles which give 52~km/s). We adopt $v\sin i=48$~km/s. We have included 18 additional measurements from Kuschnig et al. (1998), and obtain $i=45 \pm11\degr$,  $\beta=80^{+7}_{-12}\degr$ and $1560\leq B_{\rm d}\leq 2999$ (with $B_{\rm d}^{\rm min, 3,3}=1452$~G).

\subsubsection {HD 15144}

Only rather old investigations concerning the magnetic field  are published for this star. Our observations confirm the existence of a rather strong negative magnetic field at the surface of HD 15144 as already inferred by Babcock (1958) and Bonsack (1981).  Fig. 10 shows the LSD Stokes profiles when the longitudinal magnetic field was measured as  $B_\ell =-631\pm 15$~G.

We have employed the $2.99787$~d photometric period of van Genderen (1971), and we adopt $v\sin i=13$~km/s from the LSD profiles. We obtain $i=24 \pm8\degr$,  $\beta=9^{+6}_{-3}\degr$ and $2007\leq B_{\rm d}\leq 2281$ (with $B_{\rm d}^{\rm min, 3,3}=1983$~G). 

\subsubsection {21 Per, HD 18296}

21 Per was not detected significantly by Borra \& Landstreet (1980). Its rather weak longitudinal field was measured by Glagolevskij et al. (1995). We detect it easily on two occasions, and our error bars on $B_{\ell}$ are about 10 times smaller than those of previous measurements. Fig. 2 shows the LSD spectrum when the field was $213\pm 20$~G, while Fig. 3 corresponds to a cross-over phase when the longitudinal field is as small as $-39\pm 24$~G.

Although some photographic longitudinal field measurements are reported in the catalogue of Bychkov et al. (2003), their uncertainties appear to be seriously underestimated. Given the small number of precise magnetic field measurements of this star, we are unable to derive a dipole field model. However, from the peak value of the longitudinal field obtained here, we can infer that $B_{\rm d}>571$~G with $2\sigma$ confidence.

\subsubsection {9 Tau, HD 22374}

9 Tau appears to have been only observed for magnetic field by Babcock (1958). Our two observations provide an easy Zeeman detection. Fig. 11 shows the LSD Stokes profiles when the longitudinal magnetic field was measured to be $B_\ell= 523\pm 24$~G. Although we are unable to model the dipole field, we infer that $B_{\rm d}>1568$~G with $2\sigma$ confidence.

\subsubsection {56 Tau, HD 27309}

Neither Borra \& Landstreet (1980) nor Glagolevskij (as reported in the catalogue of Bychkov et al. 2003) could detect the field of this star. 56 Tau is generously detected in each of our 12 observations and the longitudinal field reaches $B_\ell=-775\pm 40$~G (Fig. 12).

The rotational period $P=1.5688840\pm 0.00000470$~d is from North \& Adelman (1995). Although Royer et al. (2002) report $v\sin i=44$~km/s, we find that this value seriously underestimates the width of the LSD profiles, and adopt $v\sin i=57$~km/s. Using these values, we derive $i=53 \pm17\degr$,  $\beta=5^{+11}_{-5}\degr$ and $2325\leq B_{\rm d}\leq 8022$~G (with $B_{\rm d}^{\rm min, 3,3}=2323$~G).  

\subsubsection {11 Ori, HD 32549}

Borra \& Landstreet (1980) could not detect the magnetic field of 11 Ori. We detect significant Stokes $V$ signatures in 4 of our 11 observations.  Our best circular polarisation detection is obtained at a crossover phase, when the longitudinal field is as weak as $-8\pm 26$~G (Fig. 13). Using the photometric period of 4.6394 d (van Leeuwen et al. 1997) and the LSD-measured $v\sin i$ of 47 km/s, we obtained $i=65\pm27\degr$,  $\beta=77^{+11}_{-74}\degr$ and $312\leq B_{\rm d}\leq 28176$~G (with $B_{\rm d}^{\rm min, 3,3}=356$~G).

\subsubsection {HD 32650}

This star is classified as B9pSi by Cowley \& Cowley (1969) and spectrophotometry was obtained by Adelman (1997, 1980). It was found to be a large amplitude variable (up to 0.1 mag in $U$-band). We did not detect its field with MuSiCoS, although we observed it several times. However the magnetic field of HD 32650 was detected (marginally) during science demonstration observations with NARVAL, and has since been detected definitively and repeatedly with this instrument. Fig. 4 shows our MuSiCoS observation on 12 February 2004. Fig. 5 shows one definite Zeeman detection with NARVAL on 12 March 07 when $B_\ell=91\pm 18$~G. 

Using the rotational period of HD 32650 from Adelman (1997) and $v\sin i=30$~km/s (from the LSD profiles) we modeled the combined MuSiCoS + NARVAL longitudinal field variation, obtaining $i=37 \pm 12\degr$,  $\beta=45^{+31}_{-30}\degr$ and  $153\leq B_{\rm d}\leq 477$ (with $B_{\rm d}^{\rm min, 3,3}=237$~G). This is one of the weakest dipole intensities inferred in this study.

\subsubsection {HD 37687}
This star is located in the direction of the Ori OB1 association.  Morrel \& Levato (1991) did not find it to be a member. However the Hipparcos parallax and proper motions do not appear conclusive alone, and radial velocity measurements will be necessary to conclude definitively on the membership. Magnetic splitting of the very sharp spectral lines of HD 37687 was not resolved in the investigation of Mathys \& Lanz (1992). We detected easily the magnetic field of this star for the very first time (Fig. 14) from only 2 LSD profiles. We find that the longitudinal magnetic field reaches $B_\ell= 766\pm 119$~G (on 17 Feb 05). From the maximum longitudinal field measured, we infer a lower limit of $B_{\rm d}>1742$~G with $2\sigma$ confidence.

\subsubsection {137 Tau, HD 39317}
The magnetic field of this star was not detected by Borra \& Landstreet (1980). In this study 137 Tau was observed 8 times and its magnetic field detected 3 times, reaching a peak value $B_\ell=222\pm 58$~G on 10 Feb 05 (Fig. 15). Using $P=2.6541$~d (Renson \& Manfroid 1981) and the LSD-measured $v\sin i$ of 45~km/s, we obtain $i=36 \pm12\degr$,  $\beta=20^{+69}_{-20}\degr$ and $113\leq B_{\rm d}\leq 2252$~G (with $B_{\rm d}^{\rm min, 3,3}=323$~G).

\subsubsection {HD 40711}
El'kin et al. (2003) observed this star and obtained a lower limit for the longitudinal magnetic field of more than several hundred G. We obtain three observations of this star, Fig. 16  shows our Stokes $V$ and Stokes $I$ LSD profiles on 13 Feb 04 when the longitudinal magnetic field was $B_\ell = -528\pm 38$~G. We obtain a $2\sigma$ lower limit on the dipole intensity of $B_{\rm d}>1492$~G.

\subsubsection {HD 43819}

This star is well studied for photometric and abundance properties (Lopez-Garcia \& Adelman 1994; Adelman \& Young 2005). Adelman \& Young (2005) found a period of 15.0313 days, stable over a 30 year time span, which is compatible with the Hipparcos period (van Leeuwen et al, 1997). Gollnow (1971) reported one insignificant longitudinal field measurement, and Bychkov et al. (2003) reported additional observations.  Fig. 17 corresponds to a crossover phase when $B_\ell$ was as weak as $-12\pm 29$~G.

Using the period of Adelman \& Young (2005) and $v\sin i=10$~km/s from the LSD profiles, and including the better-quality magnetic field measurements reported by Bychkov et al. (2003), we obtain $i=63\pm66$, $\beta=42^{+47}_{-42}\degr$ and $2626\leq B_{\rm d}\leq 78367$~G (with $B_{\rm d}^{\rm min, 3,3}=1907$~G).

\subsubsection {15 Cnc, HD 68351}

This star was previously observed by Babcock (1958) and Bohlender et al. (1993), who did not obtain any significant magnetic field detection. We observed it 16 times and detected the magnetic field 5 times. Fig. 18 presents our Stokes profiles obtained on 07 Feb 03 when the longitudinal magnetic field was $B_\ell = 325\pm 45$~G.

The period is not very well known, and we adopt that of Stepie\'n (1968; $P=4.116$~d). With $v\sin i=33$~km/s from the LSD profiles, we obtain $i=28 \pm37\degr$, $\beta=46^{+36}_{-41}\degr$ and $437\leq B_{\rm d}\leq 71486$~G (with $B_{\rm d}^{\rm min, 3,3}=762$~G).

\subsubsection {3 Hya, HD 72968}

This star has been observed by Babcock (1958) and van den Heuvel (1971) who did not obtain a significant  detection of its magnetic field. We detected it easily 13 times.  Fig. 19 presents our Stokes profiles when the longitudinal magnetic field was $B_\ell = 390\pm 16$~G.

Adelman (1998) reports a photometric period of 11.305 days, which according to our data appears to be twice the magnetic period. With a period of 5.6525~d and $v\sin i=16$~km/s, we obtain $i=61 \pm18\degr$, $\beta=5^{+7}_{-4}\degr$ and $1451\leq B_{\rm d}\leq 6702$~G (with $B_{\rm d}^{\rm min, 3,3}=1304$~G). 

\subsubsection {45 Leo, HD 90569}

This star was observed by Babcock (1958) and Bonsack (1976) who detected the magnetic field. However, their measurements have poor precision. Leroy (1993) measured linear polarisation in the direction of this star, but it is likely at least partly due to interstellar polarisation. We detected the magnetic field of 45 Leo easily 10 times. Fig. 20  presents our Stokes profiles when the longitudinal magnetic field was $B_\ell = 541\pm 23$~G.

Adopting the period $P=1.04404$~d of Adelman (2006) and measuring $v\sin i=13$~km/s, we find $i=9 \pm 4\degr$, $\beta=81^{+5}_{-7}\degr$ and $2946\leq B_{\rm d}\leq 11284$~G (with $B_{\rm d}^{\rm min, 3,3}=1634$~G).

\subsubsection {HD 94427}

This star is poorly studied, but we detected its magnetic field clearly on 8 occasions.  Fig. 21 presents our Stokes profiles when the longitudinal magnetic field was $B_\ell = 356\pm 41$~G. 

We have determined the rotational period of 1.9625~d from the LSD profile variations. With $v\sin i=8$~km/s, we obtain $i=8\pm 4\degr$,  $\beta=89^{+1}_{-4}\degr$ and $3806\leq B_{\rm d}\leq 25519$~G (with $B_{\rm d}^{\rm min, 3,3}=904$~G).

\subsubsection {EP UMa, HD 96707}

Several studies of EP UMa measured  a rather large magnetic field: van den Heuvel (1971), Bychkov et al. (1997), Leone \& Catanzaro (2001). This star was observed previously with MuSiCoS by Johnson (2004), who obtained 5 detections of Stokes $V$ signatures from 7 observations, but with a longitudinal field never larger than $119\pm 58$~G. From the total of 21 LSD Stokes $V$ profiles of EP UMa (the 7 observations of Johnson (1994) plus an additional 14 observations), we have obtained 7 definite Zeeman detections, 7 marginal detections and 7 non-detections, as defined in section 2.2. We confirm the result of Johnson (2004) - the magnetic field of EP UMa is very weak, and the longitudinal component is never larger than about 100~G.

Fig. 22 presents our LSD profiles of EP UMa when the longitudinal magnetic field was
$B_\ell = 59\pm 33$~G. The period of rotation of EP UMa has been subject of debate, with the main uncertainty being whether to choose the 3.516 d photometric period of Adelman et al  (1999), or the magnetic period of twice this value (7.0286 d) reported by Leone \& Catanzaro (2001). Using our Stokes profiles, we find independently a magnetic period of $P=3.515^{+0.009}_{-0.026}$, consistent with the photometric period. Adopting therefore the photometric period and $v\sin i=37$~km/s, we obtain $i=53 \pm16\degr$,  $\beta=90^{+0}_{-90}\degr$, and $0\leq B_{\rm d}\leq 492$~G (with $B_{\rm d}^{\rm min, 3,3}=128$~G). EP UMa exhibits one of the weakest dipole fields of any star in the sample, and it is the only star in our sample for which the longitudinal field is not significantly detected (although the presence of a field is confirmed via the Stokes $V$ profiles).

\subsubsection {65 UMa, HD 103498}

This star is part of a double system. Its magnetic field has been reported by Bychkov et al. (2003), who indicate that it reaches an extreme value of about -800~G. We observed this star 14 times and detected its magnetic field 13 times.  We find that the maximum unsigned longitudinal field is never larger than about 200~G. Fig. 23 presents our LSD profiles when the longitudinal magnetic field was $B_\ell=-166\pm 20$~G. Using the LSD Stokes $I$ and $V$ profiles, we have determined a rotational period of 15.830~d and $v\sin i=13$~km/s. The magnetic field geometry is described by $i=75 \pm 68\degr$, $\beta=80^{+10}_{-11}\degr$ and $572\leq B_{\rm d}\leq 6751$~G (with $B_{\rm d}^{\rm min, 3,3}=432$~G).

\subsubsection {21 Com, HD 108945}

21 Com is a member of the Mel 111 cluster in Coma Ber (Bounatiro, 1993). This star was observed by Borra \& Landstreet (1980) who obtained no detection of its magnetic field. Shorlin et al. (2002) detected it with MuSiCoS.  Using FORS1 at VLT, Hubrig et al. (2006) presented one measurement of the longitudinal magnetic field as a new detection, $B_\ell = -347\pm 51$~G, whereas Kochukhov \& Bagnulo (2006), using the same observational data, could not detect the magnetic field, with $B_\ell = 65\pm 116$~G. We observed 21 Com 13 times with MuSiCoS and detected its magnetic field each time.  Fig. 24 presents our LSD profiles when the longitudinal magnetic field was $B_\ell= -238\pm 55$~G.

We adopt the best magnetic period near that of Kreidl et al. (1990):  $P=2.01011$~d, and $v\sin i=65$~km/s from Shorlin et al. (2002). Our derived magnetic field geometry is $i=57 \pm18\degr$, $\beta=85^{+3}_{-61}\degr$ and $333\leq B_{\rm d}\leq 1509$~G (with $B_{\rm d}^{\rm min, 3,3}=416$~G).

\subsubsection {$\omega$ Her, HD 148112}

Borra \& Landstreet (1980) detected the magnetic field in $\omega$ Her and report  9 measurements, but with rather large uncertainties ($\sigma$ between 115 and 285 G). Shorlin et al. (2002) and Bychkov et al. (2005) add one measurement each, the values of which suggest a small amplitude of variation. Kochukhov \& Bagnulo (2006) and Hubrig et al. (2006) did not detect the magnetic field. From our 12 observations of $\omega$ Her we obtained 11 definite detections. Fig. 25 presents our LSD profiles when the longitudinal magnetic field was $B_\ell=-204\pm 21$~G.

We have adopted the rotational period $P=3.04296$~d and $v\sin i=44.5$~km/s of Hatzes (1991) (we obtained 46~km/s from the LSD profiles). The derived field geometry is $i=56 \pm16\degr$,  $\beta=3^{+11}_{-3}\degr$ and $579\leq B_{\rm d}\leq 2370$~G (with $B_{\rm d}^{\rm min, 3,3}=535$~G). 

\subsubsection {45 Her, HD 151525}

This star is the one for which detection of the magnetic field required the longest time. Babcock (1958) observed it, and Kochukhov \& Bagnulo (2006) and Hubrig et al. (2006) did not detect the magnetic field. We observed it 14 times and obtained 2 definite Zeeman detections and 3 marginal detections.  Fig. 26 presents our LSD profiles when the longitudinal magnetic field was $B_\ell = 146\pm 38$~G. 

We used the period $P=4.1164$~d and $v\sin i=35$~km/s of Hatzes (1991) (we obtained 36~km/s from the LSD profiles) to model the magnetic field geometry, obtaining $i=37\pm19\degr$, $\beta=78^{+11}_{-43}\degr$ and  $208\leq B_{\rm d}\leq 1927$~G (with $B_{\rm d}^{\rm min, 3,3}=231$~G).

\subsubsection {HD 171586}

HD 171586 is a classical Ap star (Babcock 1958), which is located in the direction of the open cluster IC 4756. However, Gray \& Corbally (2002) found that HD 171586 is not a cluster member. Only one magnetic observation by Babcock (1958) has been reported (Bychkov et al, 2003). We observed and detected this star 5 times. Fig. 27 shows the LSD profiles when the longitudinal magnetic field was measured as  $B_\ell=-375\pm 56$~G.

We have adopted the period of Winzer (1974) of $P=2.1308$~d which produces a reasonable phasing of the data, along with $v\sin i=37$~km/s from the LSD profiles. We obtain $i=48 \pm19\degr$, $\beta=46^{+40}_{-40}\degr$ and $716\leq B_{\rm d}\leq 4413$~G (with $B_{\rm d}^{\rm min, 3,3}=868$~G).

\subsubsection {HD 171782}
This star is located in the direction of IC 4556, but Herzog \& Sanders (1975) discarded it as a member because of relative proper motion measurements. No magnetic field measurements has been reported for HD 171782. We observed this star 6 times and detected it 3 times. Fig. 28 shows the LSD profiles when the longitudinal magnetic field was measured to be $B_\ell=-133\pm 78$~G.

Adopting the 4.4674~d period of Adelman \& Meadows (2002) and $v\sin i=24$~km/s measured from the LSD profiles, we obtain $i=51 \pm51\degr$, $\beta=5^{+85}_{-5}\degr$ and $B_{\rm d}\geq 213$~G (with $B_{\rm d}^{\rm min, 3,3}=584$~G). Note, due to the large uncertainties associated with $i$ and $\beta$, that the derived $B_{\rm d}$ upper limit for this star is unrealistically large.

\subsubsection {19 Lyr, HD 179527}

The magnetic field of 19 Lyr was not detected by Borra \& Landstreet (1980). We observed this star 11 times with MuSiCoS and obtained 10 Zeeman detections. Fig. 29 shows the LSD profiles at a cross-over phase, when the longitudinal magnetic field was as weak as  $B_\ell=-42\pm 42$~G. 

Using the photometric period $P=7.098$~d of Adelman \& Rice (1999) and LSD-derived $v\sin i=33$~km/s, we obtain $i=74 \pm32$, $\beta=81^{+8}_{-44}\degr$ and $409 \leq B_{\rm d}\leq 1233$ (with $B_{\rm d}^{\rm min, 3,3}=211$~G).

\subsubsection {4 Cyg, HD 183056}

The magnetic field of 4 Cyg was not detected by Borra \& Landstreet (1980). We observed the star 13 times with MuSiCoS and obtained a Zeeman detection each time.  Fig. 30 shows the LSD profiles when the longitudinal magnetic field was $B_\ell=330\pm 83$~G.

Using the LSD profile variations of this star, we obtain a rotational period of 2.9919~d and $v\sin i=26$~km/s. The derived magnetic geometry is $i=74 \pm 8\degr$, $\beta=49^{+37}_{-32}\degr$ and $1172\leq B_{\rm d}\leq 3938$~G (with $B_{\rm d}^{\rm min, 3,3}=680$~G).

\subsubsection {HD 204411}

The surface magnetic field of HD 204411 was estimated by Preston (1971) to be about 500~G. Johnson (2004) detected the magnetic field with MuSiCoS. Here we have obtained 12 observations of this star with 12 definite Zeeman detections, although the longitudinal field is always smaller than 100 G.  Fig. 31  shows the LSD profiles when the longitudinal magnetic field was $B_\ell=-88\pm 14$~G. 

We have derived a rotational period of 4.8456~d from LSD profiles. Fig. 1 shows that $B_\ell$ presents a clear variation, although HD 204411 has been for a long time considered as a very long-period photometric and/or spectroscopic variable (Preston 1970, Adelman 2003).  The spectral lines of this star are sufficiently sharp that $v\sin i$ cannot be accurate determined from our spectra, and we adopt $v\sin i=5.4$~km/s (Ryabchikova et al. 2005b). The derived magnetic geometry is $i=7 \pm 5\degr$, $\beta=81^{+7}_{-12}\degr$ and $416\leq B_{\rm d}\leq 4509$~G (with $B_{\rm d}^{\rm min, 3,3}=198$~G). This range of field intensities is consistent with the estimate of the surface field reported by Ryabchikova et al. (2005b).

\subsubsection {$\kappa$ Psc, HD 220825}

The magnetic field of $\kappa$ Psc was not detected by Borra \& Landstreet (1980). We observed the star 12 times, detecting it each time.  Fig. 32 shows the LSD profiles when the longitudinal magnetic field was $B_\ell=350\pm 24$~G.

We adopt the 1.420~d period of Kreidl \& Schneider (1989; the only published period that fits out measurements) and $v\sin i=38$~km/s from the LSD profiles. We derive $i=35 \pm54\degr$,  $\beta=83^{+7}_{-80}\degr$ and $1141\leq B_{\rm d}\leq 21045$~G (with $B_{\rm d}^{\rm min, 3,3}=865$~G).

\section{Modeling results}

\subsection{Qualitative summary of the survey}

Tables 1 and 3, Figs. 2 to 5 and Figs. 7 to 32 show that for all of the 28 weak-field Ap/Bp sample stars, the magnetic field of every sample star could be detected when sufficient tenacity and precision were employed. Thus our first conclusion is that all magnetic Ap/Bp stars, when confidently identified through spectroscopic and photometric observations, can be shown to host a measurable surface magnetic field. In every case, it was found that the longitudinal field measurements could be modeled assuming an oblique, centred dipole without any significant inconsistencies. 

\subsection {Distribution of dipole intensities $B_{\rm d}$ of the weakest-field Ap stars}

 \begin{figure}[t] 
   \centering
   \includegraphics[width=5cm, angle=-90]{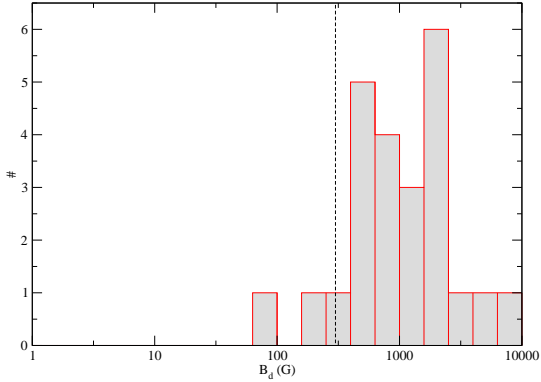}
           \label{}
\caption{Histogram of best-fit derived dipole field strengths of the stars in our sample.}
   \end{figure}

For 24 of our sample stars, a sufficient number of longitudinal field measurements were obtained to allow modeling of the dipolar magnetic field geometry in the manner described in Sect. 2.7. The results of this modeling are summarised in Table 2.

Fig. 6 shows the histogram of the derived best-fit dipole strengths $B_{\rm d}$. The remarkable characteristic of this diagram is the near-absence of Ap stars in our sample with $B_{\rm d}$ below 300~G. Only 2 stars (HD~32650 and HD~96707) are found to have best-fit model dipole field intensities below 300~G. This number increases to only 3 stars if we consider the lower limit of the dipole strength (either $B_{\rm d}^{\rm min}$ or $B_{\rm d}^{\rm min, 3.3}$, whichever is larger). Moreover, no star is constrained to have $B_{\rm d}$ below 300~G (i.e. the maximum-field model for every star has $B_{\rm d}$ larger than 300~G). 

This result cannot be due to a detection threshold effect, because the magnetic field has been detected for every star in the sample. Nor is it likely that it is due to a selection effect, as our sample was constructed specifically to include the Ap/Bp stars with the weakest magnetic fields.

What is clearly demonstrated is that the number of Ap/Bp stars does not continue to increase monotonically to arbitrarily small field strengths. Instead, it appears to plateau around 1~kG, and to decline rapidly below about 300~G.  

That this result is not an artifact of the modeling technique that we have employed is supported by the distribution of peak longitudinal fields measured for the sample, and the minimum dipole strengths derived directly from those measurements (columns 11 and 13 of Table 1). Because these minimum strengths are dependent on essentially no $i$, $\beta$ or period (dipole field strength and known limb-darkening are the only model assumptions), they provide firm $2\sigma$ lower limits on the surface field intensity of each star. In addition, those stars for which the data quality is sufficiently good show excellent agreement between the 2$\sigma$ minimum field intensities $B_{\rm d}^{\rm min}$ and $B_{\rm d}^{\rm min, 3.3}$. 

The derived surface field intensities $B_{\rm d}$ are probably themselves only lower limits on the true surface field. When sufficient data are available, detailed models of magnetic Ap stars nearly always show evidence of higher-order multipolar contributions to the magnetic field (e.g. Landstreet 1988, Landstreet \& Mathys 2000). These higher-order field components contribute only weakly to the longitudinal field variation, although they can have surface intensities comparable to that of the dipole component. Therefore the magnetic field models derived here probably systematically underestimate the true surface field intensity.

A straightforward interpretation of the behaviour observed in Fig. 6 is that there exists a minimum magnetic field strength necessary for the generation of the chemical peculiarities of Ap/Bp stars. One possible implication of this hypothesis is that there may exist a potentially large population of A-type stars which host magnetic fields which are not sufficiently strong to generate Ap-type peculiarities (Vauclair \& Vauclair 1982). However, large, high precision investigations of non-Ap tepid stars (e.g. Shorlin et al. 2002, Wade et al. 2006b, Bagnulo et al. 2006), which are able to rule out the presence of dipole fields with intensities below 300~G in many stars, suggest that no such population exists, and that the Ap/Bp stars are the only A-type stars with detectable magnetic fields.

\section{Interpretation}

The 100{\%} Zeeman detection rate obtained in our survey strongly suggests that all Ap/Bp stars host detectable magnetic fields. Moreover, it appears that a threshold magnetic field of about 300 G exists below which fields are very rare, and perhaps altogether absent.

A possible interpretation of this result is that there exists a critical field strength above which stable magnetic configurations exist
and below which any large scale field configuration is destroyed by some instability.
The instability is expected to result in the presence of opposite polarities at small length scales, thus
strongly reducing the magnitude of the integrated longitudinal field through cancellation effects.
For a sample of stars containing both stable and unstable field configurations,
this scenario would imply a strong jump in the measured values of the longitudinal fields
or a lower bound of the magnetic field, depending on the detection limit.

The existence of stable large scale magnetic fields in stars is primarily
supported by the observations of the magnetic fields of Ap stars and white dwarfs.
Theoretically, although no stable field configuration is known in an analytical form, it has been proposed that the
combination of azimuthal and poloidal field might be stable as
a recent numerical simulation tends to confirm (Braithwaite \& Spruit, 2004).
However, when the magnetic field is sufficiently weak to be wound up by differential rotation, the resulting field, predominantly azimuthal with respect to the rotation axis, can be subject to various instabilities. As recently reviewed by Spruit (1999) (see also Spruit, 2004), the most vigorous
of these instabilities is a pinch-type instability
first considered in a stellar context
by Tayler (1973).
Here, we estimate the critical magnetic field below which the winding-up process induces an instability
and above which the action of magnetic torques on the differential rotation limits the winding-up before the instability sets in.
The winding-up time scale is $t_w = 1/ (q \Omega )$ where
$q = r \| \nabla \Omega \| / \Omega = r / \ell $ is a dimensionless measure of the differential rotation. The winding-up of the axisymmetric part of the original poloidal field $\vec{B}_{\rm p}^{\rm sym}$ by the differential rotation being governed by
${\partial}_t B_{\phi} = r \sin \theta \vec{B}_{\rm p}^{\rm sym} \cdot \vec{\nabla} \Omega $, the time scale $t_w$ corresponds more specifically to the time necessary
to produce an azimuthal field component $B_{\phi}$ of the same amplitude as $\vec{B}_{\rm p}^{\rm sym}$.
On the other hand, Lorentz forces will affect the differential rotation after a Alfv\'en travel time
calculated on the shear length scale $\ell$, that is  $t_A = \ell  (4 \pi \varrho )^{1/2} / B$.
Equating both time scales gives a local  order of magnitude estimate of the critical magnetic field, $B_c \simeq (4 \pi \varrho)^{1/2} r \Omega $. Its value
can be expressed in terms of the equipartition field of a typical Ap star as follows:
\begin{equation}
\frac{B_c}{B_{eq}} \simeq 2 \left( \frac{P_{rot}}{5_{day}} \right)^{-1} \left( \frac{r}{3 R_\odot} \right) \left( \frac{T}{10^4 K} \right)^{-1/2}
\end{equation}
As $B_{eq} = 170$ G at the surface ($\tau_{5000} = 2/3$) of a typical Ap star  ($\log g=4, T_{eff} = 10^4 K$)
 the derived critical field is close to the 300 G observational threshold ($B^2_{eq}= 8 \pi P$, $P$ is the pressure). Calculation of the critical field $B_c$ for each star in the sample also 
shows that most stars satisfy $B_d > B_c $, and all stars
satisfy $B_d^{max} > B_c$. Moreover, the magnetic field of the 5 stars 
with rotational period under 2d (HD 10589, HD 27309, HD 90569, HD 94427, HD 220825) is compatible
with the dependence of $B_c$ on $P_{rot}$ $(B_c \propto P_{{rot}}^{-1})$ since their 
dipolar fields $B_d =2- 9$ kG are substantially
greater than the sample median $B_d \approx 1.4$ kG.
It is however important to stress that although the local order of 
magnitude estimate of $B_c$ is coherent with the present
observational data, a detailed and non-local modeling is required to 
specify the critical field strength below which
differential rotation destabilizes large scale field configurations.
Note that the threshold value of the magnetic field is also higher than the magnetic field threshold necessary to trigger the Tayler instability, according to the criterion given
by Spruit (1999).
The large scale field is then destroyed by the development of the instability.
An example of the non-linear evolution of such unstable configurations has been recently considered in a solar context (Brun \& Zahn 2006) confirming
that the resulting field is structured on small latitudinal length scales.

The above scenario can thus qualitatively explain the existence of an apparent lower bound in the strength of magnetic fields of Ap stars. 
By extension, such a model could provide a basis to explain why magnetic fields are observed in only a small fraction of intermediate-mass stars. If the initial magnetic field strength probability distribution of intermediate-mass stars increases (say) exponentially toward weak fields, the large majority of A-type stars, after formation, would have fields weaker than the critical field described by Eq. (8). The fields of such stars would be unstable and decay; they would therefore appear at the main sequence showing no evidence of a magnetic field.   

Another advantage of the scenario described above is that it may also naturally explain the even greater rarity of magnetic fields detected in massive stars. For a typical main sequence A0p star (with $P=5$~d, $R=3~R_\odot$ and $T_{\rm eff}=10000$~K), Eq. (8) yields $B_c\simeq 2\, B_{eq}\sim 300$~G. However, for a main sequence B0p star (with $T_{\rm eff}=31000$~K, $R=7.2~R_\odot$ and $P=2$~d), $B_c\simeq 7 B_{eq}\sim 2$~kG. With a substantially larger critical field strength, massive stars are substantially less likely to retain their magnetic fields (assuming an initial field probability distribution similar to that of Ap stars, decreasing toward strong fields). Massive stars in which relatively weak magnetic fields have been detected are all quantitatively in agreement with Eq. (8): $\beta$~Cep (Donati et al. 2001; B0, $B_c\simeq 361$~G, observed $B_{\rm d}=360$~G), $\tau$~Sco (Donati et al. 2006b; B0, $B_c\simeq 70$~G, observed $B\sim 500$~G), $\theta^1$~Ori C (Donati et al. 2002; O7, $B_c\simeq 300$~G, observed $B_{\rm d}\simeq 1.1$~kG), HD~191612 (Donati et al. 2006a, $B_c\simeq 20$~G, estimated minimum $B_{\rm d}\sim 1.5$~kG).

\section{Conclusion}

We have investigated a sample of 28 well-known spectroscopically-identified magnetic Ap/Bp stars, obtaining 282 new Stokes $V$ Zeeman signatures and longitudinal magnetic field measurements using the MuSiCoS spectropolarimeter. Magnetic field is detected in all sample stars, and the inferred longitudinal fields are significantly greater than some tens of G. To characterise the surface magnetic field intensities of the sample, we modeled the longitudinal field data to infer the intensity of the dipolar field component.  The distribution of derived dipole strengths for these stars exhibits a plateau at about 1 kG, falling off to larger and smaller field strengths. Remarkably, in this sample of stars selected for their presumably weak magnetic fields, we find only 2 stars for which the derived dipole strength is weaker than 300~G. Interpreting this ``magnetic threshold'' as a critical value necessary for the stability of large-scale magnetic fields leads to a natural explanation of the small fraction of intermediate-mass magnetic stars. It may also explain the near-absence of magnetic fields in more massive B and O-type stars.

\begin{acknowledgements}
 This research has made use of databases operated by CDS, Strasbourg, France, Vienna Atomic Line Database (VALD), Austria, and  Observatoire de Gen\`eve, Switzerland. GAW and JDL acknowledge Discovery Grant support from the Natural Sciences and Engineering Research Council of Canada (NSERC). 
We are very grateful to Dr. Victor Bychkov for providing his collected magnetic field measurements and to Dr. Thierry Roudier for providing IDL routines and making the final LSD figures.
\end{acknowledgements}

\clearpage

 \begin{figure}[t]
   \centering
   \includegraphics[width=7cm, angle=0]{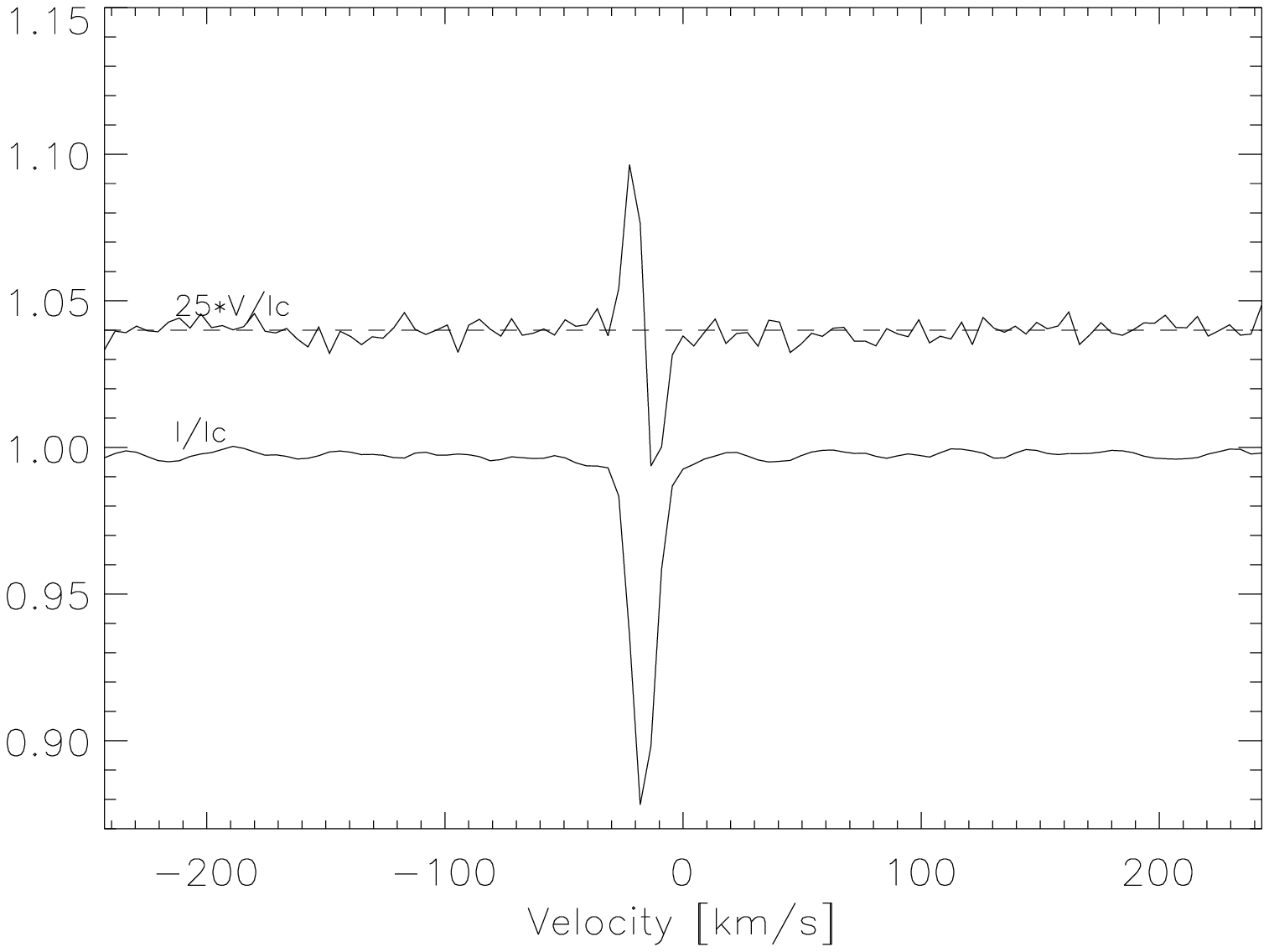}
         \label{}
\caption{LSD profiles of HN And (HD 8441) on 25 Jan 06. From bottom to top, Stokes $I$ and Stokes $V$ are presented. For display purposes, the profiles are shifted vertically, and the  Stokes $V$ profile is expanded by a factor of 25. The dashed line illustrates the zero level for the Stokes $V$ profile.}
   \end{figure}

 \begin{figure}[t]
   \centering
   \includegraphics[width=7cm, angle=0]{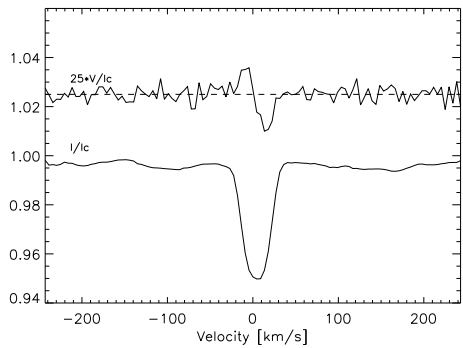}
           \label{}
\caption{LSD profiles of 43 Cas (HD 10221) on 14 Jan 06 (same as Fig. 7). }
   \end{figure}

 \begin{figure}[t]
   \centering
   \includegraphics[width=7cm, angle=0]{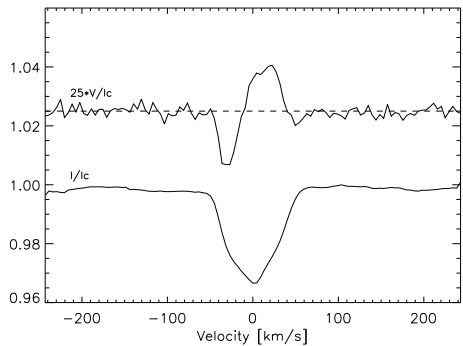}
         \label{}
\caption{LSD profiles of $\iota$ Cas (HD 15089) on 24 Aug 04 (same as Fig. 7).}
   \end{figure}

 \begin{figure}[t]
   \centering
   \includegraphics[width=7cm, angle=0]{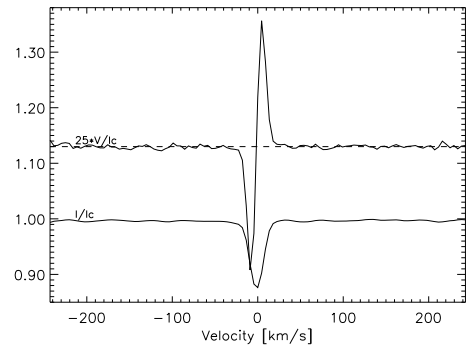}
         \label{}
\caption{LSD profiles of HD 15144 on 10 Dec 01 (same as Fig. 7). }
   \end{figure}

 \begin{figure}[t]
   \centering
   \includegraphics[width=7cm, angle=0]{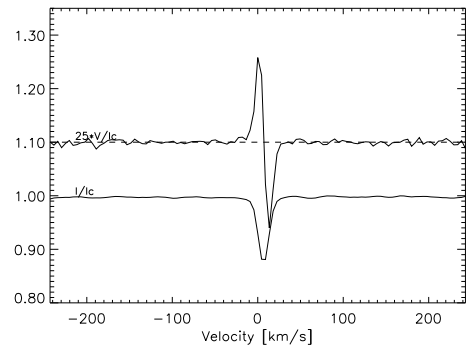}
         \label{}
\caption{LSD profiles of 9 Tau (HD 22374) on 02 Feb 04 (same as Fig. 7).}
   \end{figure}

 \begin{figure}[t]
   \centering
   \includegraphics[width=7cm, angle=0]{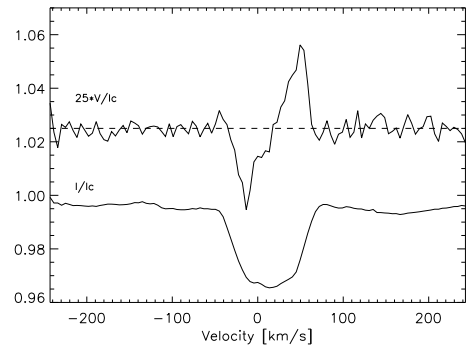}
         \label{}
\caption{LSD profiles of 56 Tau (HD 27309) on 11 Dec 01 (same as Fig. 7).}
   \end{figure}

\clearpage

 \begin{figure}[t]
   \centering
   \includegraphics[width=7cm, angle=0]{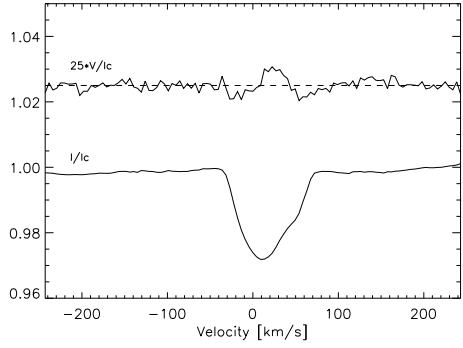}
         \label{}
\caption{LSD profiles of 11 Ori (HD 32549) on 20 Jan 06 (same as Fig. 7).}
   \end{figure}

\begin{figure}[t]
   \centering
   \includegraphics[width=7cm, angle=0]{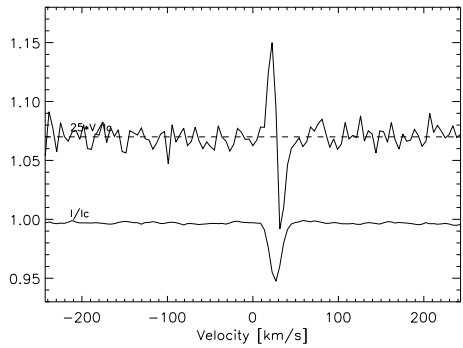}
         \label{}
\caption{LSD profiles of HD 37687 on 17 Feb 04 (same as Fig. 7).}
   \end{figure}

 \begin{figure}[t]
   \centering
   \includegraphics[width=7cm, angle=0]{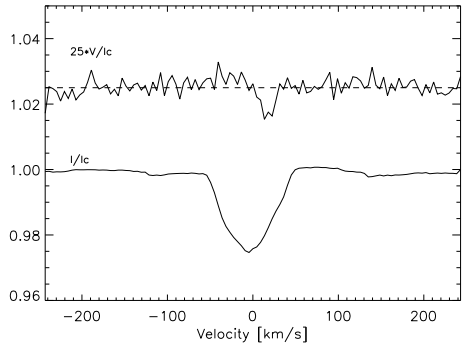}
         \label{}
\caption{LSD profiles of 137 Tau (HD 39317) on 10 Feb 05 (same as Fig. 7).}
   \end{figure}

 \begin{figure}[t]
   \centering
   \includegraphics[width=7cm, angle=0]{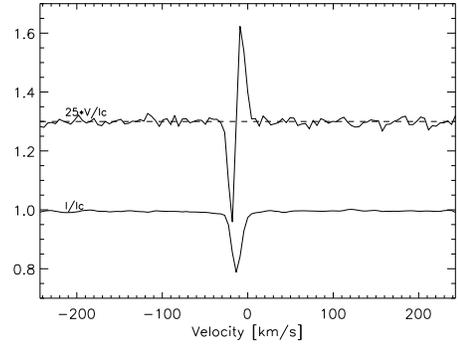}
         \label{}
\caption{LSD profiles of HD 40711 on 13 Feb 04 (same as Fig. 7).}
   \end{figure}

 \begin{figure}[t]
   \centering
   \includegraphics[width=7cm, angle=0]{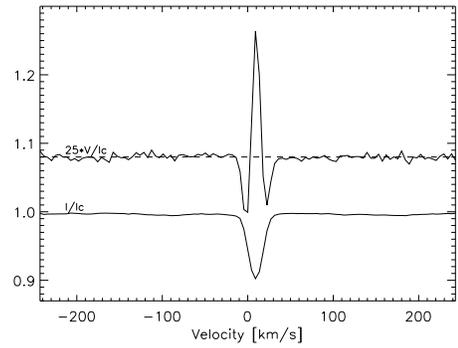}
         \label{}
\caption{LSD profiles of HD 43819 on 17 Dec 01 (same as Fig. 7).}
   \end{figure}

 \begin{figure}[t]
   \centering
   \includegraphics[width=7cm, angle=0]{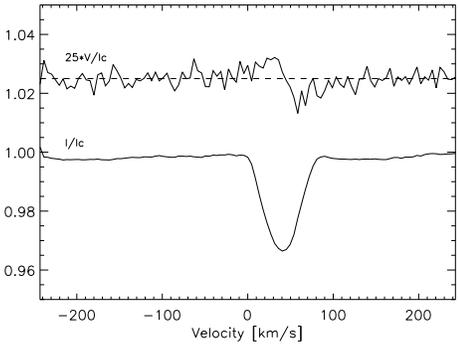}
         \label{}
\caption{LSD profiles of 15 Cnc (HD 68351) on 07 Feb 03 (same as Fig. 7).}
   \end{figure}

\clearpage

\begin{figure}[t]
   \centering
   \includegraphics[width=7cm, angle=0]{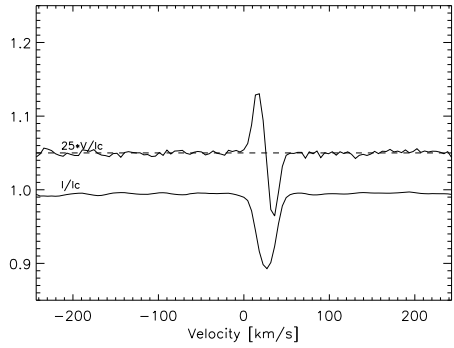}
         \label{}
\caption{LSD profiles of 3 Hya (HD 72968) on 18 Dec 01 (same as Fig. 7).}
   \end{figure}

 \begin{figure}[t]
   \centering
   \includegraphics[width=7cm, angle=0]{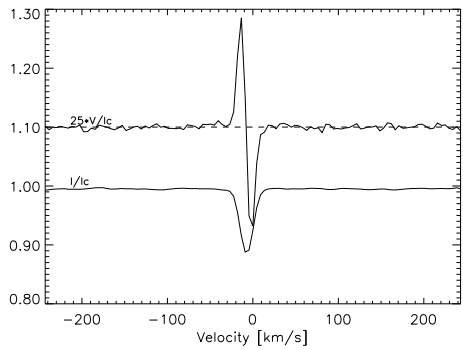}
         \label{}
\caption{LSD spectrum of 45 Leo (HD 90569) on 17 Dec 01 (same as Fig. 7).}
   \end{figure}

 \begin{figure}[t]
   \centering
   \includegraphics[width=7cm, angle=0]{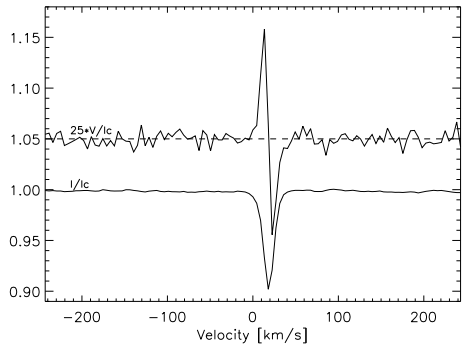}
         \label{}
\caption{LSD profiles of HD 94427 on 11 Feb 04 (same as Fig. 7).}
   \end{figure}

 \begin{figure}[t]
   \centering
   \includegraphics[width=7cm, angle=0]{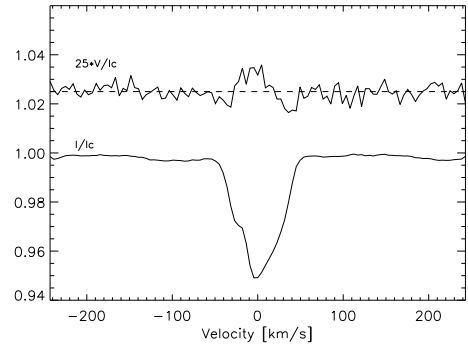}
         \label{}
\caption{LSD profiles of EP Uma (HD 96707) on 06 Feb 06 (same as Fig. 7).}
   \end{figure}

 \begin{figure}[t]
   \centering
   \includegraphics[width=7cm, angle=0]{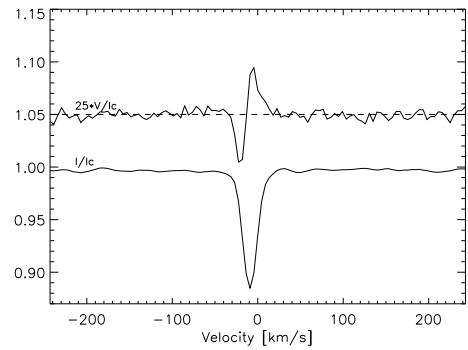}
         \label{}
\caption{LSD profiles of 65 Uma (HD 103498) on 11 Dec 01 (same as Fig. 7).}
   \end{figure}

 \begin{figure}[t]
   \centering
   \includegraphics[width=7cm, angle=0]{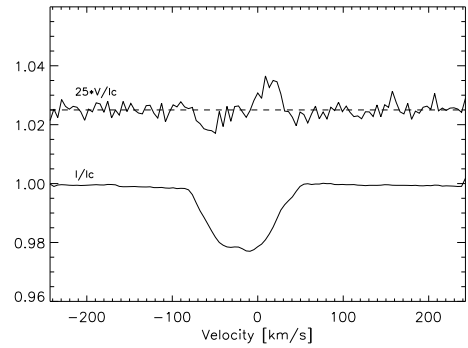}
         \label{}
\caption{LSD profiles of 21 Com (HD 108945) on 07 Feb 03 (same as Fig. 7).}
   \end{figure}

\clearpage

 \begin{figure}[t]
   \centering
   \includegraphics[width=7cm, angle=0]{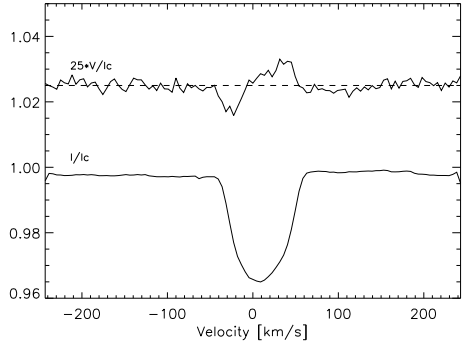}
         \label{}
\caption{LSD profiles of $\omega$ Her (HD 148112) on 03 Jul 01 (same as Fig. 7).}
   \end{figure}

 \begin{figure}[t]
   \centering
   \includegraphics[width=7cm, angle=0]{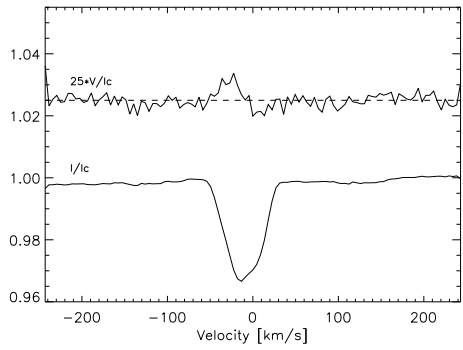}
         \label{}
\caption{LSD profiles of 45 Her (HD 151525) on 29 Jul 03 (same as Fig. 7).}
   \end{figure}

\begin{figure}[t]
   \centering
   \includegraphics[width=7cm, angle=0]{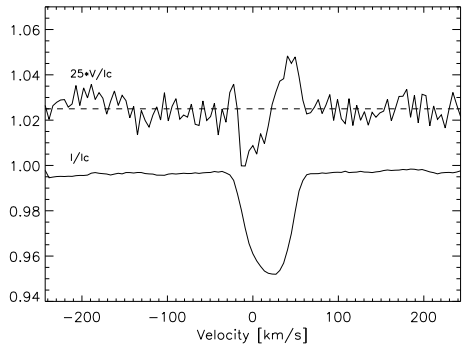}
         \label{}
\caption{LSD profiles of HD 171586 on 31 Jul 04 (same as Fig. 7).}
   \end{figure}

 \begin{figure}[t]
   \centering
   \includegraphics[width=7cm, angle=0]{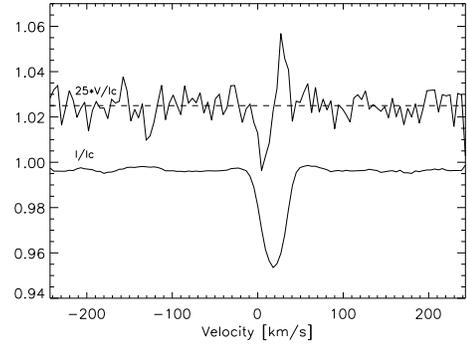}
         \label{}
\caption{LSD profiles of HD 171782 on 30 Aug 05 (same as Fig. 7).}
   \end{figure}

 \begin{figure}[t]
   \centering
   \includegraphics[width=7cm, angle=0]{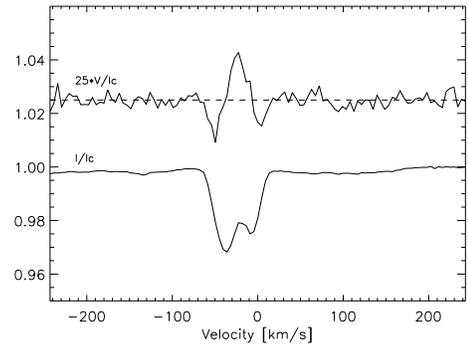}
         \label{}
\caption{LSD profiles of 19 Lyr (HD 179527) on 13 Aug 05 (same as Fig. 7).}
   \end{figure}

\begin{figure}[t]
   \centering
   \includegraphics[width=7cm, angle=0]{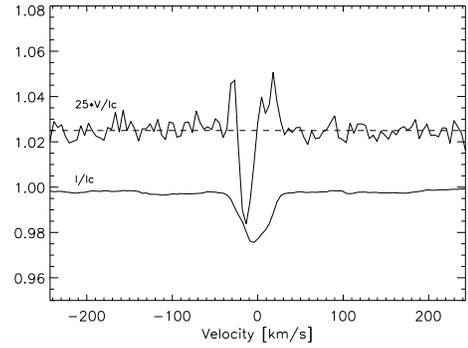}
         \label{}
\caption{LSD profiles of 4 Cyg (HD 183056) on 24 Jun 03 (same as Fig. 7).}
   \end{figure}

\clearpage

 \begin{figure}[t]
   \centering
   \includegraphics[width=7cm, angle=0]{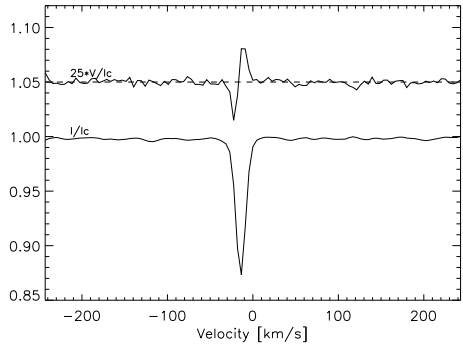}
         \label{}
\caption{LSD profiles of HD 204411 on 27 May 06 (same as Fig. 7).}
   \end{figure}

 \begin{figure}[t]
   \centering
   \includegraphics[width=7cm, angle=0]{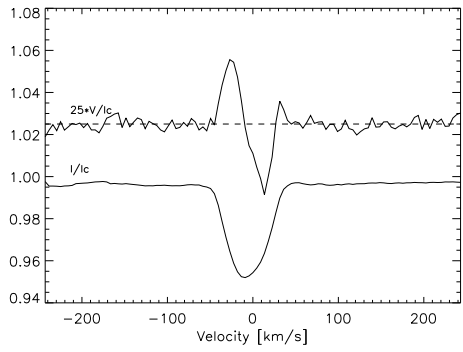}
         \label{}
\caption{LSD profiles of $\kappa$ Psc (HD 220825)on 14 Jul 04 (same as Fig. 7).}
   \end{figure}

\clearpage
\begin{longtable}{lcccccccc}
\caption{Dates, magnetic field Detection and Longitudinal magnetic field measurements. Columns give Name and HD numbers, Date, Julian Date, Signal to Noise, $\chi^2$, Detection,  longitudinal field in G, error in G. }\\
\hline\hline
  Name  & HD    & Date    &   Julian Date & S / N &  $\chi^2$ &  Detection &  $B_{l} $  (G)& $\sigma$ (G)  \\
\hline
\endfirsthead
\caption{continued.}\\
\hline\hline
  Name  & HD    &  Date   &   Julian Date & S / N &  $\chi^2$ &  Detection &  $B_{l} $  (G)& $\sigma$ (G)  \\
\hline
\endhead
\hline
\endfoot
HN And& 8441      & 12 Jul 03& 2452838.65& 220& 3.93& DD& 61& 21 \\
& & 30 Jul 04& 2453217.64& 180& 2.16& DD& -20& 31 \\
& & 31 Aug 04& 2453249.66& 180& 1.92& DD& 40& 30 \\
& & 25 Jan 06& 2453761.35& 220& 7.62& DD& 157& 18 \\
& & 31 Jan 06& 2453767.28& 210& 4.34& DD& 74& 23 \\
& & 02 Feb 06& 2453769.29& 300& 3.98& DD& 88& 18 \\
& & 04 Feb 06& 2453771.28& 190& 2.81& DD& -4& 27 \\
& & 05 Feb 06& 2453772.26& 240& 3.53& DD& 28& 18 \\
43 Cas& 10221& 01 Dec 01& 2452245.40& 310& 3.05& DD& 89& 38 \\
& &11 Dec 01& 2452255.42& 310& 1.52& DD& 67& 34 \\
& &12 Dec 01& 2452256.37& 420& 4.06& DD& 87& 32 \\
& & 16 Dec 01& 2452260.37& 410& 0.82& nd& 30& 29 \\
& & 06 Feb 04& 2453042.29& 270& 0.85& nd& -41& 48 \\
& & 12 Feb 04& 2453048.29& 310& 2.35& DD& 49& 43 \\
& & 10 Jan 06& 2453746.37& 440& 3.90& DD& 121& 30 \\
& & 12 Jan 06& 2453748.37& 420& 2.35& DD& 57& 30\\
& & 14 Jan 06& 2453750.37& 320& 2.42& DD& 148& 34 \\
& & 21 Jan 06& 2453757.34& 450& 4.07& DD& 77& 31\\
$\iota$ Cas& 15089& 04 Feb 04& 2453040.34& 180& 1.91& DD& 182& 125\\
& & 06 Feb 04& 2453042.33& 420& 6.18& DD&-222& 38\\
& & 24 Aug 04& 2453242.60& 450& 6.35& DD&-348& 28\\
& & 29 Aug 04& 2453247.65& 360& 5.41& DD&-152& 45\\
& & 03 Aug 05& 2453586.54& 480& 6.52& DD& 448& 26\\
& & 04 Aug 05& 2453587.63& 550& 13.89& DD& -42& 25\\
& & 06 Aug 05& 2453589.65& 600& 11.80& DD& 389& 20\\
& & 07 Aug 05& 2453590.56& 630& 9.51& DD& -298& 24\\
& & 08 Aug 06& 2453591.59& 540& 8.24& DD& 486& 23\\
& & 11 Aug 05& 2453594.55& 460& 10.74& DD& -142& 27\\
& & 24 Aug 05& 2453607.55& 420& 5.14& DD& 433& 37\\
& & 02 Sep 05& 2453616.51& 620& 8.77& DD& 8& 29\\
&15144 & 09 Dec 01& 2452253.38& 290& 32.23& DD& -568& 15\\
& & 10 Dec 01& 2452254.40& 340& 38.95& DD& -631& 15\\
& & 08 Feb 05& 2453410.32& 260& 17.71& DD& -558& 24\\
& & 30 Aug 05& 2453613.57& 300& 31.24& DD& -558& 15\\
& & 01 Sep 05& 2453615.55& 420& 44.51& DD& -623& 13\\
& & 03 Sep 05& 2453617.56& 360& 38.14& DD& -580& 14\\
21 Per& 18296& 01 Sep 05& 2453615.64& 680& 8.01& DD& 213& 20\\
& & 02 Sep 05& 2453616.59& 630& 12.27& DD& -39& 24\\
9 Tau & 22374 & 02 Feb 04& 2453038.30& 190& 18.51& DD& 523& 24\\
& & 23 Aug 05& 2453606.61& 250& 19.28& DD& -195& 18\\
56 Tau & 27309& 01 Dec 01& 2452245.51& 310& 3.01& DD& -693& 72\\  
& & 11 Dec 01& 2452255.51& 370& 5.19& DD& -615& 49\\
& & 17 Dec 01& 2452261.47& 400& 5.78& DD& -768& 56\\
& & 06 Jan 02& 2452281.42& 420& 4.90& DD& -681& 50\\
& & 12 Jan 03& 2452652.41& 280& 4.41& DD& -717& 61\\
& & 24 Aug 04& 2453242.66& 210& 2.52& DD& -654& 86\\
& & 31 Aug 04& 2453249.63& 260& 2.80& DD& -714& 90\\
& & 16 Jan 03& 2452656.40& 330& 3.58& DD& -771& 68\\
& & 11 Aug 05& 2453594.66& 460& 4.85& DD& -625& 46\\
& & 13 Aug 05& 2453596.67& 440& 7.17& DD& -804& 50\\
& & 14 Aug 05& 2453597.67& 530& 6.81& DD& -775& 40\\
& & 03 Sep 05& 2453617.64& 490& 5.15& DD& -574& 39\\
11 Ori& 32549    & 02 Dec 01& 2452246.52& 610& 1.44& md& 172& 46 \\
& & 08 Dec 01& 2452252.40& 520& 1.27& nd& 30& 43 \\
& & 12 Dec 01& 2452256.73& 630& 1.21& nd& 171& 37 \\
& & 05 Feb 04& 2453041.45& 500& 1.37& md& 44& 42 \\
& & 17 Feb 04& 2453053.46& 440& 0.97& nd& 11& 69 \\
& & 11 Jan 06& 2453747.57& 200& 0.87& nd& -47& 104 \\
& & 12 Jan 06& 2453748.53& 600& 1.26& nd& -150& 37 \\
& & 16 Jan 06& 2453752.56& 550& 1.49& md& -186& 39 \\
& & 20 Jan 06& 2453756.41& 780& 2.39& DD& -8& 26 \\
& & 21 Jan 06& 2453757.39& 590& 1.22& nd& -96& 33 \\
& & 22 Jan 06& 2453758.52& 520& 1.19& nd& 66& 57 \\
& 32650& 09 Dec 01& 2452253.55& 370& 1.15& nd& 55& 57 \\
& & 10 Dec 01& 2452254.54& 400& 0.84& nd& 47& 69 \\
& & 05 Feb 04& 2453041.48& 390& 0.95& nd& 73& 59 \\
& & 12 Feb 04& 2453048.46& 420& 0.87& nd& 98& 54 \\
& & 10 Jan 06& 2453746.56& 500& 1.00& nd& 105& 39 \\
& & 13 Jan 06& 2453749.58& 510& 1.14& nd& 105& 43 \\
& & 19 Jan 06& 2453755.53& 390& 1.14& nd& 122& 67 \\
& & 21 Jan 06& 2453757.50& 480& 1.07& nd& 72& 42 \\
& & 22 Jan 06& 2453758.40& 430& 0.76& nd& -42& 63 \\
& & 27 Nov 06& 2454067.48& 562& 1.12& nd& 29& 27\\
& & 29 Nov 06& 2454069.57& 734& 1.42& md& 47& 16\\
& & 30 Nov 06& 2454070.54& 866& 1.04& nd& 4& 18\\
& & 14 Dec 06& 2454084.55& 684& 0.83& nd& 3& 23\\
& & 11 Mar 07& 2454171.34 &  1188  & 1.73& DD & 0    &  14 \\          
& & 12 Mar 07& 2454172.39 &   918  & 2.24& DD & 91   &   18\\          
& & 13 Mar 07& 2454173.34 &   1439 & 2.37& DD & 39   &   10\\          
& & 14 Mar 07& 2454174.38 &   1070 & 0.74& nd & 24   &   16\\          
& & 15 Mar 07& 2454175.36 &   862  & 1.78& md & 47   &   19\\   
&37687 & 17 Feb 04& 2453053.42& 130& 4.57& DD& 565& 108\\
& & 17 Feb 05& 2453419.45& 120& 5.63& DD& 766& 119\\
137 Tau& 39317 & 01 Dec 01& 2452245.55& 330& 1.11& nd& 68& 70\\
& & 12 Dec 01& 2452256.52& 410& 1.96& DD& 200& 53\\
& & 10 Feb 05& 2453412.47& 490& 1.60& DD& 222& 58\\
& & 23 Jan 06& 2453759.39& 360& 1.16& nd& 199& 66\\
& & 24 Jan 06& 2453760.41& 330& 1.48& md& 129& 73\\
& & 31 Jan 06& 2453767.45& 360& 0.92& nd& -18& 68\\
& & 01 Feb 06& 2453768.38& 390& 1.84& DD& 94& 58\\
& & 03 Feb 06& 2453770.45& 460& 1.38& nd& 59& 52\\
& 40711& 13 Feb 04& 2453049.44& 83& 13.38& DD& -528& 38\\
& & 08 Feb 05& 2453410.48& 100& 1.64& md& -97& 71\\
& & 09 Feb 05& 2453411.50& 44& 1.29& nd& 40& 184\\
&43819 & 10 Dec 01&  2452254.64& 230& 22.04& DD& 631& 31\\
& & 11 Dec 01& 2452255.54& 180& 16.61& DD& 642& 34\\
& & 13 Dec 01& 2452257.75& 150& 14.66& DD& 724& 43\\
& & 16 Dec 01& 2452260.61& 230& 21.16& DD& 93& 30\\
& & 17 Dec 01& 2452261.60& 240& 15.69& DD& -12& 29\\
& & 06 Jan 02& 2452281.56& 200& 17.04& DD& 74& 35\\
& & 04 Feb 06& 2453771.47& 230& 21.50& DD& 534& 36\\
& & 05 Feb 06& 2453772.46& 340& 29.00& DD& 628& 25\\
15 Cnc& 68351& 08 Dec 01& 2452252.64& 350& 1.44& md& 148& 46\\
& & 11 Dec 01& 2452255.63& 370& 1.38& md& 70& 47\\
& & 07 Feb 03& 2452678.46& 340& 2.079& DD& 276& 41\\
& & 08 Feb 03& 2452679.54& 320& 0.98& nd& 124& 62\\
& & 29 Jan 04& 2453034.59& 180& 0.92& nd& -16& 102\\
& & 13 Feb 04& 2453049.48& 250& 1.09& nd& 117& 75\\
& & 14 Feb 04& 2453050.54& 300& 0.86& nd& 58& 61\\
& & 23 Jan 06& 2453759.55& 350& 1.16& nd& 88& 46\\
& & 24 Jan 06& 2453760.57& 300& 1.29& nd& 137& 57\\
& & 25 Jan 06& 2453761.55& 400& 1.06& nd& 80& 47\\
& & 29 Jan 06& 2453765.50& 210& 1.24& nd& 306& 69\\
& & 31 Jan 06& 2453767.50& 370& 1.13& nd& 72& 51\\
& & 01 Feb 06& 2453768.50& 400& 1.29& nd& 146& 42\\
& & 02 Feb 06& 2453769.47& 410& 1.53& md& 140& 37\\
& & 03 Feb 06& 2453770.50& 490& 1.34& md& 84& 39\\
& & 07 Feb 06& 2453774.48& 490& 1.15& nd& 62& 35\\
3 Hya &72968 & 10 Dec 01& 2452254.69& 300& 7.73& DD& 283&19\\
& & 13 Dec 01& 2452257.60& 200& 9.34& DD& 379& 27\\
& & 15 Dec 01& 2452259.61& 200& 8.58& DD& 359& 28\\
& & 16 Dec 01& 2452260.65& 240& 6.59& DD& 300& 23\\
& & 17 Dec 01& 2452261.64& 330& 15.51& DD& 338& 17\\
& & 18 Dec 01& 2452262.65& 340& 16.24& DD& 390& 16\\
& & 05 Jan 02& 2452280.58& 280& 12.60& DD& 375& 20\\
& & 06 Jan 02& 2452281.59& 280& 12.97& DD& 378& 19\\
& & 12 Jan 06& 2453748.58& 330& 16.49& DD& 389& 16\\
& & 13 Jan 06& 2453749.54& 360& 17.12& DD& 427& 16\\
& & 19 Jan 06& 2453755.42& 150& 6.99& DD& 430& 35\\
& & 20 Jan 06& 2453756.52& 370& 16.84& DD& 376& 15\\
& & 04 Feb 06& 2453771.51& 310& 15.06& DD& 390& 18\\
45 Leo& 90569& 07 Dec 01&  2452251.65& 320& 29.78& DD& 430& 23\\ 
& & 13 Dec 01& 2452257.64& 150& 14.57& DD& 311& 43\\
& & 15 Dec 01& 2452259.65& 160& 10.64& DD& 71& 40\\
& & 16 Dec 01& 2452260.68& 260& 20.29& DD& 171& 24\\
& & 17 Dec 01& 2452261.68& 260& 22.203& DD& 541& 23\\
& & 18 Dec 01& 2452262.68& 310& 25.83& DD& 154& 22\\
& & 05 Jan 02& 2452280.65& 240& 24.13& DD& 438& 27\\
& & 06 Jan 02& 2452281.63& 300& 26.87& DD& 382& 27\\
& & 03 Feb 06& 2453770.63& 270& 22.89& DD& 465& 27\\
& & 05 Feb 06& 2453772.53& 290& 23.29& DD& 195& 26\\
& 94427& 07 Feb 03& 2452678.55& 120& 8.71& DD& 277& 40\\  
& & 21 Jan 04& 2453026.65& 64& 3.81& DD& -534& 94\\
& & 11 Feb 04& 2453047.58& 110& 7.26& DD& 356& 41\\
& & 09 May 06& 2453865.36& 150& 2.75& DD& -56& 33\\
& & 18 May 06& 2453874.39& 100& 5.79& DD& -358& 55\\
& & 24 May 06& 2453880.38& 120& 8.02& DD& -386& 45\\
& & 28 May 06& 2453884.36& 110& 6.76& DD& -403& 57\\
& & 01 Jun 06& 2453888.37& 79& 4.72& DD& -304& 82\\
EP Uma &96707 & 10 Dec 01& 2452254.74& 290& 1.51& md& 10& 38 \\
& & 11 Dec 01& 2452255.67& 320& 1.67& DD& -2& 31 \\
& & 15 Dec 01& 2452259.68& 180& 0.85& nd& 50& 66 \\
& & 11 Jan 03& 2452650.66& 240& 1.14& nd& 41& 44 \\
& & 16 Jan 03& 2452655.57& 270& 1.41& md& 3& 44 \\
& & 19 Jan 03& 2452658.68& 180& 1.44& md& 24& 63 \\
& & 29 Jan 04& 2453034.65& 170& 0.92& nd& -33& 65 \\
& & 01 Feb 04& 2453037.67& 160& 1.23& nd& 145& 62 \\
& & 02 Feb 04& 2453038.63& 170& 1.45& md& 13& 75 \\
& & 05 Feb 04& 2453041.64& 280& 1.37& md& 30& 38 \\
& & 08 Feb 04& 2453044.64& 250& 1.44& md& -15& 40 \\
& & 13 Feb 04& 2453049.63& 310& 2.16& DD& -48& 33 \\
& & 17 Feb 04& 2453053.62& 260& 1.58& md& -17& 40 \\
& & 23 Jan 06& 2453759.59& 280& 1.64& DD& -23& 40 \\
& & 24 Jan 06& 2453760.61& 260& 0.87& nd& -2& 47 \\
& & 25 Jan 06& 2453761.59& 350& 2.03& DD& -16& 28 \\
& & 31 Jan 06& 2453767.58& 340& 0.97& nd& -42& 33 \\
& & 05 Feb 06& 2453772.62& 330& 1.62& DD& 28& 32 \\
& & 06 Feb 06& 2453773.50& 340& 1.81& DD& 59& 33 \\
& & 07 Feb 06& 2453774.55& 400& 1.12& nd& -18& 28 \\
& & 08 Feb 06& 2453775.49& 330& 1.73& DD& -29& 29 \\
65 Uma& 103498& 11 Dec 01& 2452255.73& 200& 6.28& DD& -166& 20\\
& & 13 Dec 01& 2452257.69& 100& 2.97& DD& -231& 43\\
& & 15 Dec 01& 2452259.72& 100& 2.87& DD& -142& 44\\
& & 15 Dec 01& 2452259.75& 100& 2.26& DD& -112& 45\\
& & 16 Dec 01& 2452260.72& 170& 4.44& DD& 16& 24\\
& & 17 Dec 01& 2452261.71& 150& 4.54& DD& 88& 26\\
& & 18 Dec 01& 2452262.72& 150& 5.67& DD& 183& 27\\
& & 06 Jan 02& 2452281.67& 170& 6.41& DD& 176& 25\\
& & 06 Feb 05& 2453408.58& 57& 1.67& MD& 388& 143\\
& & 06 Feb 05& 2453408.60& 89& 1.18& ND& 202& 153\\
& & 08 Feb 05& 2453410.62& 190& 4.54& DD& 33& 28\\
& & 10 Feb 05& 2453412.65& 150& 3.45& DD& -163& 42\\ 
& & 02 Feb 06& 2453769.60& 240& 9.01& DD& 169& 19\\
& & 05 Feb 06& 2453772.57& 220& 7.72& DD& -73& 20\\
21 Com& 108945   & 13 Dec 01& 2452257.73& 160& 1.46& md& -130& 139\\
& & 16 Dec 01& 2452260.75& 320& 2.64& DD& -131& 55\\
& & 17 Dec 01& 2452261.75& 330& 2.73& DD& 75& 62\\
& & 18 Dec 01& 2452262.75& 340& 2.79& DD& -108& 53\\
& & 06 Jan 02& 2452281.70& 290& 1.67& DD& 204& 60\\
& & 16 Jan 03& 2452656.63& 380& 2.51& DD& -199& 48\\
& & 07 Feb 03& 2452678.63& 340& 2.29& DD& -238& 55\\
& & 03 Feb 05& 2453405.67& 420& 2.64& DD& -24& 55\\
& & 08 Feb 05& 2453410.66& 440& 2.19& DD& 75& 58\\
& & 08 Feb 05& 2453410.69& 270& 1.50& DD& 80& 102\\
& & 16 Jan 06& 2453752.66& 320& 2.14& DD& 174& 59\\
& & 19 Jan 06& 2453755.57& 310& 1.85& DD& 15& 84\\
& & 20 Jan 06& 2453756.58& 440& 3.66& DD& -70& 43\\
$\omega$ Her & 148112& 03 Jul 01& 2452094.40& 700& 3.60& DD& -204& 21\\ 
& & 28 Jul 01& 2452119.41& 330& 1.75& DD& -211& 44\\
& & 16 Jul 03& 2452837.39& 620& 2.96& DD& -181& 22\\
& & 20 Jul 03& 2452841.39& 420& 1.58& DD& -218& 44\\
& & 04 Aug 03& 2452851.38& 590& 2.07& DD& -183& 31\\
& & 03 Aug 05& 2453586.41& 400& 2.07& DD& -217& 34\\
& & 05 Aug 05& 2453588.39& 710& 3.11& DD& -175& 20\\
& & 06 Aug 05& 2453589.39& 310& 1.63& DD& -233& 45\\
& & 07 Aug 05& 2453590.40& 700& 2.38& DD& -172& 24\\
& & 08 Aug 05& 2453591.39& 270& 1.21& nd& -182& 73\\
& & 15 Aug 05& 2453598.38& 670& 3.35& DD& -187& 20\\
& & 04 Sep 05& 2453618.36& 590& 2.42& DD& -149& 26\\
45 Her &151525 & 07 Jul 01& 2452098.41& 340& 0.85& nd& -199& 136 \\
& & 28 Jun 03& 2452819.41& 460& 1.60& DD& -79& 40 \\
& & 30 Jun 03& 2452821.39& 180& 0.99& nd& 66& 96 \\
& & 07 Jul 03& 2452828.42& 300& 0.77& nd& -63& 90 \\
& & 11 Jul 03& 2452832.40& 450& 1.31& nd& -59& 51 \\
& & 21 Jul 03& 2452842.38& 380& 1.25& nd& 43& 49 \\
& & 29 Jul 03& 2452850.37& 470& 1.81& DD& 146& 38 \\
& & 15 Jan 04& 2453020.70& 180& 0.63& nd& 84& 128 \\
& & 08 Feb 04& 2453044.72& 350& 1.21& nd& 44& 63 \\
& & 14 Feb 04& 2453050.71& 370& 1.35& md& 40& 58 \\
& & 11 Jul 04& 2453198.46& 260& 1.13& nd& -130& 75 \\
& & 12 Jul 04& 2453199.47& 410& 1.37& md& 98& 46 \\
& & 15 Jul 04& 2453202.48& 420& 1.23& nd& 0& 45 \\
& & 17 Jul 04& 2453204.44& 300& 1.48& md& 179& 62 \\
& 171586& 31 Jul 04& 2453218.45& 200& 3.01& DD& -375& 56\\
& & 06 Aug 04& 2453224.46& 140& 1.59& DD& -371& 93\\
& & 13 Aug 04& 2453231.41& 220& 3.84& DD& -96& 55\\
& & 01 Jun 06& 2453888.49& 200& 2.54& DD& -40& 51\\
& & 03 Jun 06& 2453890.52& 230& 2.98& DD& -39& 46\\
& 171782& 05 Aug 04& 2453223.45& 120& 1.44& md& -200& 99\\
& & 12 Aug 04& 2453230.44& 93& 0.86& nd& -290& 276\\
& & 28 Aug 05& 2453611.41& 100& 1.69& DD& -386& 118\\
& & 30 Aug 05& 2453613.37& 160& 2.57& DD& -333& 78\\
& & 01 Jun 06& 2453888.53& 80&  1.28& nd& -668& 194\\
& & 04 Jun 06& 2453891.51& 100& 1.17& nd& -214& 144\\
19 Lyr &179527   & 18 Jul 01& 2452109.42& 250& 1.92& DD& 78& 79 \\
& & 06 Jul 03& 2452827.49& 310& 1.75& DD& 156& 46 \\
& & 08 Jul 03& 2452829.51& 230& 1.46& md& -21& 75 \\
& & 09 Jul 03& 2452830.52& 270& 2.11& DD& -173& 59 \\
& & 27 Jul 03& 2452848.37& 240& 1.28& nd& 85& 92 \\
& & 04 Aug 05& 2453587.41& 340& 1.52& md& 137& 45 \\
& & 12 Aug 05& 2453595.45& 470& 2.44& DD& 113& 35 \\
& & 13 Aug 05& 2453596.39& 430& 3.78& DD& -42& 42 \\
& & 14 Aug 05& 2453597.39& 490& 3.38& DD& -127& 31 \\
& & 14 Aug 05& 2453597.43& 490& 3.22& DD& -136& 31 \\
& & 23 Aug 05& 2453606.38& 350& 2.36& DD& -21& 52 \\
4 Cyg& 183056& 24 Jul 01& 2452115.49& 270& 6.34& DD& -46& 90\\
& & 24 Jun 03& 2452815.53& 280& 5.28& DD& -341& 85\\
& & 02 Aug 05& 2453585.47& 530& 3.93& DD& 281& 47\\
& & 04 Aug 05& 2453587.44& 410& 6.30& DD& -255& 60\\
& & 05 Aug 05& 2453588.43& 490& 3.48& DD& 237& 52\\
& & 06 Aug 05& 2453589.47& 430& 8.79& DD& -265& 61\\
& & 12 Aug 05& 2453595.42& 580& 11.48& DD& -254& 46\\
& & 13 Aug 05& 2453596.51& 520& 9.14& DD& -259& 49\\
& & 14 Aug 05& 2453597.47& 550& 3.92& DD& 226& 43\\
& & 14 Aug 05& 2453597.56& 580& 4.68& DD& 268& 43\\
& & 15 Aug 05& 2453598.41& 470& 9.26& DD& -291& 55\\
& & 30 Aug 05& 2453613.41& 540& 9.34& DD& -177& 45\\
& & 01 Sep 05& 2453615.40& 580& 3.77& DD& 290& 42\\
&204411   & 04 Jul 03& 2452825.56& 340& 6.026& DD& -37& 12 \\
& & 07 Jul 03& 2452828.58& 370& 9.37& DD& -33& 14 \\
& & 21 Jul 04& 2453208.57& 210& 4.64& DD& -25& 20 \\
& & 23 Jul 04& 2453210.48& 210& 4.00& DD& -87& 19 \\
& & 08 Aug 05& 2453591.43& 210& 4.97& DD& -9& 20 \\
& & 15 Aug 05& 2453598.45& 450& 9.86& DD& -68& 9 \\
& & 24 Aug 05& 2453607.39& 320& 6.95& DD& -82& 13 \\
& & 08 May 06& 2453864.65& 260& 4.81& DD& -87& 16 \\
& & 10 May 06& 2453866.66& 310& 3.13& DD& -3& 13 \\ 
& & 24 May 06& 2453880.56& 480& 13.57& DD& -28& 9 \\
& & 27 May 06& 2453883.55& 280& 6.16& DD& -88& 14 \\
& & 06 Jun 06& 2453893.63& 420& 7.58& DD& -75& 10 \\
$\kappa$ Psc & 220825& 03 Jul 01& 2452094.60& 220& 4.85& DD& -191& 57\\
& & 24 Jun 03& 2452815.62& 280& 9.19& DD& 156& 42\\
& & 14 Jul 04& 2453201.53& 410& 7.23& DD& 350& 29\\
& & 21 Jul 04& 2453208.60& 290& 5.61& DD& 345& 35\\
& & 22 Jul 04& 2453209.61& 160& 4.64& DD& -31& 70\\
& & 02 Aug 05& 2453585.62& 480& 8.11& DD& -207& 23\\
& & 04 Aug 05& 2453587.60& 450& 15.50& DD& 141& 25\\
& & 05 Aug 05& 2453588.55& 520& 5.36& DD& -285& 22\\
& & 06 Aug 05& 2453589.62& 590& 18.08& DD& 133& 21\\
& & 07 Aug 05& 2453590.52& 550& 14.61& DD& 299& 20\\
& & 15 Aug 05& 2453598.58& 450& 6.01& DD& -317& 25\\
& & 24 Aug 05& 2453607.51& 300& 7.93& DD& 263& 38\\
\hline
\hline
\end{longtable}

\end{document}